\documentclass[twocolumn,preprintnumbers,showpacs,aps,
               amsmath,amsfonts]{revtex4}
\usepackage{epic,eepic}
\usepackage{hyperref}

\begin{document}

\newcommand{\lp}{\left}
\newcommand{\rp}{\right}
\newcommand{\be}{\begin{eqnarray}}
\newcommand{\ee}{\end{eqnarray}}
\newcommand{\beq}{\begin{equation}}
\newcommand{\eeq}{\end{equation}}
\newcommand{\ba}{\begin{array}}
\newcommand{\ea}{\end{array}}

\newcommand{\braket}[2]{{\langle{#1}|{#2}\rangle}}
\newcommand{\ket}[1]{{|{#1}\rangle}}
\newcommand{\bra}[1]{{\langle{#1}|}}
\newcommand{\av}[1]{\langle#1\rangle}
\newcommand{\bbraket}[2]{{\left\langle{#1}|{#2}\right\rangle}}
\newcommand{\bket}[1]{{\left|{#1}\right\rangle}}
\newcommand{\bbra}[1]{{\left\langle{#1}\right|}}
\newcommand{\bav}[1]{\left\langle#1\right\rangle}

\newcommand{\mypmatrix}[1]{\begin{pmatrix}#1\end{pmatrix}}
\newcommand{\Z}{\mathbb{Z}}
\newcommand{\C}{\mathbb{C}}
\newcommand{\Q}{Q}
\newcommand{\sdp}[1]{\times_{#1}}
\newcommand{\mymod}{\ \text{mod}\ }
\newcommand{\e}[1]{\text{e}^{#1}}
\newcommand{\diag}[1]{\text{diag}\left(#1\right)}
\newcommand{\Tr}[1]{\ \text{Tr}\lp(#1\rp)}


\preprint{CALT-68-2440}

\title{Anyon computers with smaller groups}

\author{Carlos Mochon}
\email{carlosm@theory.caltech.edu}
\affiliation{Institute for Quantum Information, 
California Institute of Technology,
Pasadena, CA 91125,USA}

\date{March 28, 2004}

\begin{abstract}
Anyons obtained from a finite gauge theory have a computational power
that depends on the symmetry group. The relationship between group
structure and computational power is discussed in this paper. In particular,
it is shown that anyons based on finite groups that are solvable but not 
nilpotent are capable of universal quantum computation. This extends previously
published results to groups that are smaller, and therefore more practical.
Additionally, a new universal gate-set is built out of an operation called 
a probabilistic projection, and a quasi-universal leakage correction scheme
is discussed.
\end{abstract}



\maketitle

\section{Introduction}

The two main obstacles to building a practical quantum computer are the
decoherence produced by the environment and the need for 
unitary operators of sufficiently high precision.
Topological quantum computation provides a way of encoding quantum information
in non-local observables that are protected from the environment, thereby 
solving the first problem. In some instances, the second problem
can also be addressed by using operations that only depend on topological 
invariants.

Anyons present a concrete realization of the ideas of 
topological quantum computation that may have practical implementations. 
An anyon is a particle that has exotic quantum statistics, and exists in
a two-dimensional space.
Anyons carry certain topological charges which cannot be locally measured 
or modified, and can therefore be used to store protected quantum information.
The charges can be detected, though, using two elementary operations 
called braiding and fusion. In the first operation, the positions of
two anyons in the plane are exchanged, causing their world-lines to braid. 
Because clockwise and counterclockwise rotations can be distinguished
in two dimensions, braiding can produce results more complicated
than the usual bosonic and fermionic cases. The second operation involves 
fusing two anyons into a single anyon that carries the combined charges of 
the original particles. In both cases, the results only depend on the
charges and the topological class of the paths involved. These operations
can be used as a basis of error-free gates that manipulate the stored 
information.

Many different models of anyons can be constructed by specifying 
different spectra of charges together with a set of braiding and fusion rules.
A convenient and physically inspired set of models can be obtained from
the electric and magnetic charges of a two-dimensional finite-group 
gauge theory. These models depend on a finite group $G$, which acts as
the symmetry of the gauge theory. While every finite group produces
a consistent model, the computational power of the resulting anyons
depends on the structure of the group.

Previous work by the same author \cite{me2002} has shown that finite 
non-solvable groups produce anyons capable of universal quantum computation.
However, the smallest finite non-solvable group is $A_5$, the even permutations
of five objects, which has $60$ elements. Unfortunately, anyons with a large
symmetry group are less likely to be found in nature, and are also harder to
engineer. A more desirable symmetry group would be $S_3$, with only
$6$ elements. The purpose of this paper is to study the feasibility of
quantum computation with these smaller groups. In fact, it will be shown
that the groups that are solvable but not nilpotent, which includes
$S_3$ as the smallest case, produce anyons capable of universal quantum 
computation. The caveat, though, is that the constructions in this paper
require both electric and magnetic charges, whereas magnetic charges alone
were sufficient in the non-solvable case \cite{me2002}. The use of electric
charges complicates the procedure significantly, and will occupy the bulk
of the discussion.

The ideas of this paper and its predecessor \cite{me2002} are built on the
foundations laid out by Kitaev \cite{Kitaev:1997wr}, who introduced the notion
of a quantum computer based on anyons. The first concrete description
for the group $A_5$ was done by Ogburn and Preskill
in Refs.~\cite{Ogburn:1998,Preskill:1997uk}. An unpublished construction
for the group $S_3$ was also worked out by Kitaev, and its use of 
electric charges served as a basis for much of the present work.

The organization of this paper is as follows: Section~\ref{sec:review}
contains a review of the basic properties of anyons, and describes
the notation used in this paper. The next two sections
prove the universality of anyons based on groups that are semidirect
products of certain cyclic groups of prime order, which
includes the important case of $S_3$. Section~\ref{sec:sdp}
constructs an abstract set of gates out of the fundamental anyon 
operations, whereas Section~\ref{sec:gates} proves that this gate-set is
universal. In Section~\ref{sec:gengrp}, the discussion is expanded to general
finite groups, and the relationship between group structure and computational
power is established. This section will also review the definitions of 
solvability and nilpotency. The main result of this paper, which is the 
feasibility of universal quantum computation with anyons from groups that 
are solvable but non-nilpotent, is proven in Section~\ref{sec:meat}. 
The discussion in Section~\ref{sec:meat} is motivated by
Section~\ref{sec:sdp}, and includes many of the same steps, but the 
details are significantly more complicated. Finally, Section~\ref{sec:leak}
discusses a leakage correction scheme that can be applied to anyons,
as well as many other quantum systems.

\section{\label{sec:review}Review and Notation}

In this section we present a brief review of the anyon properties and
operations that will be used throughout the paper. Notation for anyon pairs,
qudit bases, and a special type of gate will be introduced.

The gauge theory model for anyons that is used in this paper 
was first presented in  Ref.~\cite{Bais:1992pe}, and is 
summarized in the review in  Ref.~\cite{deWildPropitius:1995hk}. 
Our notation will be
closer to the one used in the author's previous work \cite{me2002}, though.
In the interest of space, we present only a very brief review of the basic
ideas and refer the reader to the above references for further details.
The only new ideas introduced in this section beyond Ref.~\cite{me2002} are
the notation for electric charge pairs and a gate called the probabilistic 
projection.

\subsection{Anyon model and spectrum}

To fully characterize a system with anyons, we must specify a set of braiding
and fusion rules. A set of consistent rules can be obtained from 
the behavior of electric and magnetic charges in a finite-group gauge theory.
Though many other anyon models exist, including models with continuous groups,
and models with a finite spectrum that cannot be obtained from a
gauge theory, only the finite-group gauge-theory model will be discussed
in this paper.

For each finite group $G$, there is a gauge theory with symmetry group $G$ 
that contains anyons.
The anyonic spectrum of the finite-group gauge theory consists of electric 
charges, magnetic charges, and particles called dyons which carry both 
electric and magnetic charge. The magnetic charges, also known as fluxes,
are labeled by elements of the group $G$. The electric charges are 
labeled by an irreducible unitary representation $R$
of $G$, and have an internal state that transforms as a vector under $R$.
The dyons are labeled by an element of $g\in G$ and a representation of
the stabilizer of $g$. The dyons, however, will not play an significant role
in this paper.

\subsection{Magnetic charge pairs}

We begin by discussing the braiding rules for the magnetic charges,
which will be the most important particles in this paper.
The basic rule for magnetic charges is the following: when two fluxes are
exchanged, the flux of one is conjugated by the flux of the other.
Though this is the basic interaction that will be used between magnetic 
charges, it has the undesirable consequence that moving single fluxes 
through the system can introduce unwanted correlations. Therefore,
it will be necessary to work with pairs of fluxes of trivial total flux.

For any $g\in G$, we define the state $\ket{g}$
to denote a magnetic charge pair, where the first anyon
has flux $g$, and the second anyon has flux $g^{-1}$. Because the 
effects of actions on the compensating flux $g^{-1}$ will mimic the effects
on the flux $g$, we will generally not mention them explicitly. In fact,
we shall refer to the state $\ket{g}$ as ``a state of flux $g$,''
by which we describe the flux of the first anyon, rather than the total
flux of the anyon pair which shall always be trivial.

A general state of $n$ magnetic charge pairs has the form
\be
\ket{\Psi} = \sum_{g_1,\dots,g_n\in G} 
\alpha_{g_1,\dots,g_n} \ket{g_1}\otimes\cdots\otimes\ket{g_n},
\ee

\noindent
where $\alpha_{g_1,\dots,g_n}$ are the complex quantum amplitudes.
Due to the existence of superselection sectors, which will be discussed below,
the $g_i$ in the above sums are restricted to a single conjugacy class, 
which may be different for each $i$.

Because we are dealing with pairs of trivial total flux, any two states
can be swapped simply by exchanging the physical position of the anyons:
\beq
\sum_{g_1,g_2\in G} 
\alpha_{g_1,g_2} \ket{g_1}\otimes\ket{g_2}
\longrightarrow 
\sum_{g_1,g_2\in G} 
\alpha_{g_1,g_2} \ket{g_2}\otimes\ket{g_1}.
\eeq

\noindent
By repeatedly exchanging pairs of adjacent anyons, any two pairs of anyons
can be exchanged. This operation will also allow us to move pre-made ancillas 
into the computational space, and to move anyons that have been fused 
out of the computational space.

The basic interaction between pairs is a pass-through operation
by which one pair passes in between a second pair. The result of the operation
leaves the first pair invariant, but conjugates the second pair by
either the flux of the first pair:
\beq
\sum_{g_1,g_2\in G} 
\alpha_{g_1,g_2} \ket{g_1}\otimes\ket{g_2}
\longrightarrow \!\!\!
\sum_{g_1,g_2\in G} 
\alpha_{g_1,g_2} \ket{g_1}\otimes\ket{g_1g_2g_1^{-1}},
\eeq

\noindent
or its inverse:
\beq
\sum_{g_1,g_2\in G} 
\alpha_{g_1,g_2} \ket{g_1}\otimes\ket{g_2}
\longrightarrow \!\!\!
\sum_{g_1,g_2\in G} 
\alpha_{g_1,g_2} \ket{g_1}\otimes\ket{g_1^{-1}g_2g_1},
\eeq

\noindent
depending on the direction of the pass-though. By using
the swap operation, the pass-though can be performed on any two pairs of
anyons. 

Furthermore, the above operation can be generalized to a conjugation by
a function of the fluxes of a set of anyons. That is, consider a function
$f:G^n\rightarrow G$ that can be written as a product of its inputs,
their inverses, and fixed elements of $G$. For example,
\be
f\lp(g_1,g_2\rp) = c_1 g_1^{-1} c_2 g_1 c_3 g_1 c_4 g_2 c_5,
\ee

\noindent
where the $\{c_i\}$ are fixed elements of $G$, and this case has $n=2$.
Then, if we assume the existence of ancillas of the form $\ket{g}$ for each
$g\in G$, we can perform the unitary transformation
\beq
\ket{g_1}\otimes\ket{g_2}\otimes\ket{g_3} \longrightarrow
\ket{g_1}\otimes\ket{g_2}\otimes\ket{f(g_1,g_2)g_3f(g_1,g_2)^{-1}},
\eeq

\noindent
where we have denoted its action on basis elements, and the general 
transformation follows by linearity. The operation is implemented
by conjugating in sequence by the entries of $f$, starting from right to left,
where an ancilla of known flux is used for every fixed element of $f$.
In general, a conjugation by a function can be performed with any
state as target, and any number of inputs, as long as the function
can be written in product form.

\subsection{Electric charge pairs}

In addition to pairs of magnetic charges, this paper will 
often deal with pairs of electric charges, where the
first charge transforms under the irreducible representation $R$, and the 
second charge transforms under the complex conjugate representation $R^*$.
Of course, for some representations $R^*\simeq R$, which will not be a problem
for what follows.

We introduce the bases $\lp\{\ket{i_R}\rp\}$ and $\lp\{\ket{j_{R^*}}\rp\}$
on which the representations act. The indices $i,j$ take values from $1$ 
to $d_R$, the dimension of the representation. 
We assume that the basis vectors are compatible in the sense that 
\be
\bra{i_{R^*}} R^*(g) \ket{j_{R^*}} = \bra{i_{R}} R(g) \ket{j_{R}}^*.
\ee

The combined state of the two charges is spanned by the vectors 
$\ket{i_R}\otimes\ket{j_{R^*}}$ and can be described by specifying a 
$d \times d$ matrix $M$
\be
\ket{M}_R \equiv \frac{1}{\sqrt{d_R}} 
                 \sum_{i,j} M_{i j} \ket{i_R}\otimes\ket{j_{R^*}},
\ee

\noindent
where we have introduced a convenient normalization factor.

We will be interested in the braiding and fusion properties of these states. 
However, when two electric charges move past each other, even when they are 
not in pairs, their charges remain unchanged. It is only 
the magnetic fluxes that have an effect on the electric charges. In 
particular, when a magnetic flux $g$ goes around an
electric charge, the flux remains invariant, but the charge transforms
as if multiplied by $g$ in the representation $R$. 
Starting with a state $\ket{M}_R$, if the flux circles the first electric 
charge, then it becomes
\be
U(g)\otimes I \ket{M}_R 
&=& \frac{1}{\sqrt{d_R}}  
    \sum_{i,j,k} R_{i k}(g) M_{k j} \ket{i_R}\otimes\ket{j_{R^*}} \nonumber\\ 
&=& \ket{R(g) M}_R
\ee

\noindent
where $R(g) M$ is the matrix obtained by left multiplying
$M$ by the element $g$ in the representation $R$. Similarly, if we act on 
the second charge, we obtain
\be
I \otimes U(g) \ket{M}_R 
&=& \frac{1}{\sqrt{d_R}} 
    \sum_{i,j,k} M_{i k} R^*_{j k}(g) \ket{i_R}\otimes\ket{j_{R^*}} \nonumber\\
&=& \frac{1}{\sqrt{d_R}} 
    \sum_{i,j,k} M_{i k} R^\dagger_{k j}(g) \ket{i_R}\otimes\ket{j_{R^*}} 
\nonumber \\ 
&=& \ket{M R(g^{-1})}_R
\ee

\noindent
where we have used the fact that $R$ is unitary.

Note that, just as in the case of the magnetic charges, if we have a function
$f(\{g_i\})$ of some anyon fluxes, written out in product form, then we
can apply this function to our charges
\be
\ket{M}_R \longrightarrow U\lp(f\rp) \otimes I  \ket{M}_R = 
\ket{R\lp(f\rp)M}_R
\ee

\noindent
by applying sequentially from right to left the elements of the product.

\subsection{Superselection sectors, fusion, and vacuum pairs}

Before describing the fusion rules for the magnetic and electric
charges, we need to address the issue of superselection sectors, which is
familiar to particle physicists. A superselection sector is a subspace of
a Hilbert space that is invariant under all the implementable transformations.
A useful analogy is to consider the Hilbert space of a particle called the
nucleon, spanned by the four states
\be
\ket{0\uparrow}, \ket{0\downarrow}, \ket{1\uparrow}, \ket{1\downarrow},
\ee

\noindent
corresponding to a spin-$\frac{1}{2}$ particle with two possible charge 
values. This is nothing more than the direct sum of the Hilbert spaces of 
the proton and the neutron:
\be
\mathcal{H}_{\mbox{nucleon}} = \mathcal{H}_{\mbox{proton}}
\oplus \mathcal{H}_{\mbox{neutron}}.
\ee

\noindent
At the energies of atomic physics, it is not possible to measure
in the proton plus neutron basis, or to perform a unitary rotation along this 
direction. Therefore, we could say that a nucleon automatically decoheres
into either a proton or a neutron.

A similar situation occurs with the anyons. Each conjugacy class of $G$ is a
magnetic charge superselection sector. The irreducible representations
are the electric charge superselection sectors. When given an unknown 
anyon---for example, an anyon created from the vacuum---we can assume that
it has decohered into a specific, though possibly unknown, conjugacy 
class and/or irreducible representation. Furthermore, when storing quantum 
information, it will be important to keep the computational space in a 
single superselection sector to avoid decoherence.

Let $\ket{\Psi}$ be a pair of anyons created from the vacuum. We may
assume that each anyon has decohered into a specific superselection sector.
Furthermore, because a vacuum pair must consist of a particle with its
antiparticle, the two superselection sectors are related. That is, the
pair must have vacuum quantum numbers and be able to fuse back into the vacuum.
Therefore, if the first anyon is a magnetic charge with flux in a given 
conjugacy class, the second anyon will be a magnetic charge with flux in the
inverse conjugacy class. If the first anyon is an electric charge of 
representation $R$, then the second anyon will be a 
electric charge of the complex conjugate representation. Finally, if one
anyon is a dyon, then so is the other.

In the case of magnetic charges, there is exactly one state with 
vacuum quantum numbers in each conjugacy class. The state is 
\be
\ket{\mbox{Vac}(\mathcal{C})} = \frac{1}{\sqrt{|\mathcal{C}|}}
\sum_{g\in\mathcal{C}} \ket{g},
\ee

\noindent
where $\mathcal{C}$ is a conjugacy class of $G$. Note that, given our
notation, the above state is an entangled state of two anyons. In the
case of electric charges, the vacuum state for representation $R$ is
simply $\ket{R(I)}_R$, where $R(I)$ is the $d_R \times d_R$ identity matrix.

The operation of fusion is in a sense the inverse of
vacuum pair creation. Fusing two anyons produces a single anyon that
must carry the total magnetic and electric charges of the pair. In the special
case when both total charges are trivial (i.e., one of the above vacuum states)
the state can fuse into the vacuum, leaving no particle behind,
and transferring its energy to some other medium such as photons.
In theory, this case can easily be detected in the laboratory, and
is the primary way of obtaining measurement results.

In the case of magnetic charges, the net resulting flux is just the product
of the two fluxes, where the ordering of the product depends on some 
conventions which will not be important here.
While one of our standard anyon pairs always has trivial total
flux, we sometimes may fuse anyons from different pairs to determine if their
flux is equal. Even if the total flux is trivial, though, the pair may not fuse
into the vacuum but may produce an electric charge. This will be the case if 
the state transforms non-trivially under simultaneous conjugation of 
both anyons.

The fusion of two electric charges can only produce another electric charge
(or the vacuum, which is the charge carrying the trivial representation).
To calculate the possible products of fusion, note that fusion implies 
that a flux can no longer be braided around only one of the two electric 
charges. Mathematically, it is a restriction to the diagonal transformations
\be
\ket{M}_R \longrightarrow U(g) \otimes U(g) \ket{M}_R = 
\ket{R(g) M R(g^{-1})}_R.
\ee

However, the above action of the group is not irreducible on this space.
The vector space spanned by all possible states $\ket{M}_R$ decomposes 
into invariant subspaces. 
The invariant subspaces  correspond to electric charges 
transforming under irreducible representations. 
The probability of obtaining each irreducible representation corresponds 
to the magnitude of the state vector projected down to the appropriate 
invariant subspace. 
Furthermore, after fusion, it is no longer possible to measure the relative 
phase between the different representations and therefore decoherence
occurs in the representation basis.

The net result of fusion is a mixed state of different 
representations. Which representations occur is determined by the 
decomposition of $R(g)\otimes R^*(g)$  into irreducible representations. 
The probability of obtaining each of these representations is determined 
by the projection of $M$ to the different invariant subspaces.

In particular, the trace of $M$ is the unique invariant under conjugation by 
$G$ (which is the content of Schur's lemma). Therefore the probability of 
fusion into the vacuum is
\be
P_{\mbox{vac}} = \lp|\braket{R(1)}{M}_R\rp|^2 = \lp|\frac{\Tr{M}}{d_R}\rp|^2.
\ee

\subsection{\label{sec:reqs}Requirements for the physical system}

To complete our review of the properties of anyons, we will list the
operations, ancillas, and measurements that we assume
are available on any realistic system, and which we will use
to build our quantum gate-set:
\begin{enumerate}
\item We can braid or exchange any two particles.
\item We can fuse a pair of anyons and detect whether there is a particle
left behind or whether they had vacuum quantum numbers.
\item We can produce a pair of anyons in a state that is chosen at
random from the two particle subspace that has vacuum quantum numbers.
\item We have a supply of ancillas of the form $\ket{g}$ for any $g\in G$.
\item We have a supply of ancillas of the form $\ket{R(I)}_R$ for any 
irreducible unitary representation $R$.
\end{enumerate}

The last two requirements are the only questionable ones, as it is not
obvious how to produce this reservoir of calibrated electric and magnetic
charges. In fact, since many of these ancillas will be destroyed during fusion,
the reservoir will have to have a large number of ancillas of each type. 

One of the main difference between the constructions in this paper,
and the one used in producing computations with non-solvable groups
\cite{me2002}, is that the latter case required no electric charge ancillas,
which may be harder to produce. Additionally, Ref.~\cite{me2002} presented a
protocol for producing the magnetic ancillas for a simple non-abelian 
group. The production of calibrated flux and charge ancillas
for the groups discussed in the present paper, though similar, will not 
be addressed here.

A final note is that the requirement of calibrated magnetic charge ancillas
will have to be slightly modified in Section~\ref{sec:newreq}, in order to work
with certain large groups.

\subsection{Notation for Qudits}

Throughout this paper it will be useful to perform computations
with qudits rather than the usual qubits. We define our computational
basis as the states $\ket{i}$ for $0\leq i < d$, where we will assume
that $d$ is prime. The unitary $Z$ and $X$ gates can be defined as follows:
\be
Z \ket{i} &=& \omega^{i} \ket{i},\\
X \ket{i} &=& \ket{i+1},
\ee

\noindent
where $\omega$ is a fixed non-trivial $d^{th}$ root of unity, and sums
are understood to be modulo $d$. 
As usual, the eigenstates of $Z$ correspond to the computational basis.
We can also introduce the eigenstates of $X$:
\be
\ket{\tilde i} = \frac{1}{\sqrt{d}} \sum_{j=0}^{d-1} \omega^{-i j} \ket{j},
\ee

\noindent
which have the following transformations under the action of our unitary
gates:
\be
Z \ket{\tilde i} &=& \ket{\widetilde{i-1}},\\
X \ket{\tilde i} &=& \omega^{i} \ket{\tilde i}.
\ee

Note that when appropriate, we shall assume all operations are modulo $d$
without further comment.

\subsection{Probabilistic projection onto $\mathcal{K}$}

To conclude with the introduction of notation, we define a new type of
gate called a probabilistic projection onto a subspace. The operation is 
essentially a projective measurement that distinguishes between a subspace
$\mathcal{K}$, and its orthogonal complement. However, the operation has a 
one-sided probability of error, corresponding to a failure to notice the 
projection into $\mathcal{K}$. 

For example, consider an operation that emits a photon if and only if 
the state is projected into the subspace $\mathcal{K}$. The photon is then 
received at a photodetector that has a probability $0<p \leq 1$ of absorbing 
the photon.
A photon will never be detected if the state was projected into the complement
of $\mathcal{K}$, but even if the measurement projected
into $\mathcal{K}$, the photodetector may remain silent.

To formalize the idea of a probabilistic projection, let $\mathcal{K}$ be a 
subspace of a Hilbert space $\mathcal{H}$, and let 
$P_{\mathcal{K}}$ be the projection onto $\mathcal{K}$. We define a 
probabilistic projection onto $\mathcal{K}$ as a two-outcome POVM
with operators 
\be
F_0 = p_{PP} P_{\mathcal{K}},\ \ \ \ F_1 = 1 - p_{PP} P_{\mathcal{K}},
\ee

\noindent
where  $0<p_{PP}\leq 1$. We say that we can do a probabilistic projection onto 
$\mathcal{K}$ if we can do the above operation for any fixed $p_{PP}$.

Furthermore, we demand that if outcome $0$ is obtained when applying the
operation to a state $\ket{\Psi}$, we obtain the state
\be
\ket{\Psi_0} = 
\frac{P_{\mathcal{K}} \ket{\Psi}}{\sqrt{\bra{\Psi}P_{\mathcal{K}}\ket{\Psi}}}.
\ee 

\noindent
On the other hand, if we get the result $1$, we will consider the state 
damaged, and trace it out of our computational system.

As an example consider
\be
\ket{\Psi} = \frac{1}{\sqrt{2}} \lp(\ket{0}\otimes\ket{1} 
                                    + \ket{1}\otimes\ket{0}\rp),
\ee

\noindent
and let $\mathcal{K} = \lp\{\ket{0}\rp\}$. Applying a probabilistic projection
to the first qubit, we obtain with probability $p_{PP}/2$ the state
\be
\ket{\Psi_0} = \ket{0}\otimes\ket{1},
\ee

\noindent
and with probability $1-p_{PP}/2$ we obtain the mixed state
\be
\rho_1 = \frac{1}{2-p_{PP}}\bigg[\lp(1-p_{PP}\rp)\ket{1}\bra{1} 
                                 + \ket{0}\bra{0}\bigg],
\ee

\noindent
where we have already traced out the first qubit. Notice that if the 
probabilistic projection onto $\ket{0}$ is applied to both qubits 
simultaneously, it is possible to obtain the result $1$ twice, but it is 
not possible to obtain the result $0$ twice.

\section{\label{sec:sdp}Base Case: $G = \Z_p \sdp{\theta} \Z_q$}

Before tackling the general case of groups that are solvable but not nilpotent,
we will describe the procedure for producing quantum computation using
a special type of group based on the semidirect product. The construction
for these groups is very similar to the general case, but can be described in
more concrete terms. In particular, these groups are very useful in 
eliminating operations whose usefulness is unclear in the general case, 
but that have no computational power when reduced to this special case.

\subsection{Algebraic structure}

We will be interested in the groups $G = \Z_p \sdp{\theta} \Z_q$, 
the semidirect
product of the cyclic groups of order $p$ and $q$. We assume that $p\neq q$
are both prime and that the function $\theta$ is non-trivial, which guarantees
that $G$ is not nilpotent.

The group can be described using two generators $a$ and $b$ which satisfy
the relations:
\be
a^p = 1, \ \ \ b^q = 1,\ \ \ \ 
b a b^{-1} = a^t,
\ee

\noindent
where specifying an integer $t$ between $0$ and $p$ is equivalent to specifying
the function $\theta:\Z_q\rightarrow \mbox{Aut}(\Z_p)$ used for the semidirect
product. We will require that $t\neq 1$ which is equivalent to $\theta$ being 
non-trivial. Furthermore, consistency requires that
\be
a = b^q a b^{-q} = a^{t^q} \ \Longrightarrow\ 
t^q = 1 \mymod p,
\ee

\noindent
which can always be solved for some $t$ as long as 
$q$ divides $p-1$. We henceforth assume that $p$, $q$, and $t$ have been 
chosen in a self-consistent fashion.

The best example of one of these groups, and in fact the smallest non-abelian 
group, is $S_3$. This group can be expressed as $\Z_3 \sdp{\theta} \Z_2$, 
with $t = 2$. We can choose $a$ to be any order three element such as 
$(123)$, and we can choose $b$ to be any order two element such as $(12)$.

The first example of such a group with odd order is $\Z_7 \sdp{\theta} \Z_3$ 
with $t = 2$ or $t = 4$, both of which are equivalent. One of the most
important features of this example is that not all the non-trivial powers
of $a$ are conjugate to one another. 
The elements $a$, $a^2$, and $a^4$ form one conjugacy
class, whereas the elements $a^3$, $a^5$, and $a^6$ form another.

Both of the above examples will be revisited when we discuss group 
representations and fusion of electric charges.

\subsection{Computational basis}

We choose a qudit computational basis
\be
\ket{i} = \ket{a^i b a^{-i}},
\ee

\noindent
for $0\leq i<p$. Note that all these states are unique because 
$a^i b a^{-i}= a^{i(1-t)} b$, and $a^{1-t}$ is a non-trivial generator
of the group $\Z_p$. We are therefore using a complete conjugacy class for
the computational subspace.

While the above choice of computational subspace may seem arbitrary,
most other choices are either equivalent or less powerful.
The conjugacy classes $a^i b^j a^{-i}$, for different non-trivial values 
of $j$, are all equivalent. 
Dyons with these fluxes are also equivalent since they are 
just the combination of the above states with electric charges 
that cannot be detected by braiding. Finally, the powers of $a$ and pure
electric charges are suboptimal as they are difficult to entangle (for more
on this see the discussion on using nilpotent groups in 
Section~\ref{sec:gengrp}).

Initializing a quantum computer in this basis is easy, as we have assumed
the existence of flux ancillas in the state $\ket{0}$, which can be used as 
computational anyons.
We therefore turn to the task of implementing gates on this space.

\subsection{\label{sec:braid}Operations involving braiding fluxes}

We begin by characterizing the operations that can be achieved
by braiding fluxes. Fix a target qudit which we will be 
conjugating, and assume that it is in the computational subspace.
We can conjugate this qudit by the fluxes of arbitrary ancillas in the group.
It can also be conjugated by the fluxes of other qudits, which we will also
assume to have a definite flux in the computational subspace (as the effect
of a superposition of fluxes can be inferred by linearity).

Let us begin with the case when only one qudit (in addition to the target) is
involved. If the source qudit is in a state $\ket{g}$, then the target will get
conjugated by an expression
\be
f(g) = c_1 g c_2 g c_3 \dots c_n,
\ee

\noindent
for some $n$, where the $\lp\{c_j\rp\}$ are fixed elements of $G$ 
corresponding to the ancillas used. 
Of course, these elements represent the product of any ancillas
that were used in series, and can equal the identity if no ancillas were used.

Because of the structure of the group, all the fixed elements can be expanded
as $c_i = a^{j_i} b^{k_i}$ for some integers $j_i$, $k_i$. Furthermore, since
the source flux is in the computational basis, it can be written out as
$g = a^x b a^{-x} = a^{x(1-t)} b$, for some $x$.
Inserting these expressions, we get
\be
f(g) = a^{j_1} b^{k_1} a^{x(1-t)} b a^{j_2} b^{k_2} \dots a^{j_n} b^{k_n}.
\ee

Using the group relation $b a^i = a^{it}b$, we can move all the $b$'s to 
the right, and combine factors to get
\be
f(g) = a^{\alpha} a^{\beta x} b^{\delta},
\ee

\noindent
for some integers $\alpha$, $\beta$ and $\delta$. The effect of each of 
these factors can be considered separately. Conjugating by $a^\alpha$ is
just the application of the gate $X^\alpha$. Conjugating by $a^{\beta x}$
is just a controlled-$X$ from the source to the target, repeated $\beta$
times. Finally, conjugating by $b$ maps $\ket{i}$ to $\ket{it}$.
This operation can be generated using a controlled-$X$ gate, and an ancilla 
$\ket{0}$:

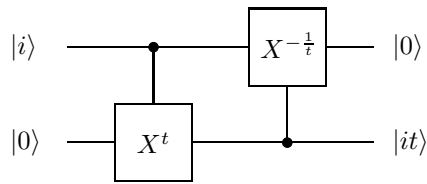
\begin{center}
\setlength{\unitlength}{1in}
\begin{picture}(3,1.1)(0,0)
{\put(.7,.8){\line( 1, 0){.95}}}
{\put(2.3,.8){\line( -1, 0){.25}}}
{\put(.7,.3){\line( 1, 0){.25}}}
{\put(2.3,.3){\line( -1, 0){.95}}}
{\put(1.15,.8){\line( 0,-1){.3}}}
{\thinlines \put(1.15,.8){\circle*{.05}}}
{\put(0.95,.1){\framebox(.4,.4){}}}
{\put(1.85,.3){\line( 0,1){.3}}}
{\thinlines \put(1.85,.3){\circle*{.05}}}
{\put(1.65,.6){\framebox(.4,.4){}}}
\put(1.15,.3){\makebox(0,0){$X^t$}}
\put(1.86,.81){\makebox(0,0){$X^{-\frac{1}{t}}$}}
\put(.4,.8){\makebox(0,0)[l]{$\ket{i}$}}
\put(.4,.3){\makebox(0,0)[l]{$\ket{0}$}}
\put(2.4,.8){\makebox(0,0)[l]{$\ket{0}$}}
\put(2.4,.3){\makebox(0,0)[l]{$\ket{it}$}}
\end{picture}
\end{center}

\noindent
where $-1/t$ is computed modulo $p$. Following the above circuit, we can 
either replace the original qudit with the ancilla, or use a swap, which 
can also be built out of controlled-$X$ gates.

So far we have shown that the $X$ and controlled-$X$ gates generate the
set of operations achieved by conjugations. However, we have yet to show
that these operations are in fact included in the set of achievable operations.
The $X$ gate is rather trivial as it is a conjugation by an ancilla of flux 
$a$. The controlled-$X$ is a conjugation by the function 
\be 
f(g)&=&\lp( g b^{-1} \rp)^{1/(1-t) \mymod p} \nonumber\\
&=& \lp( a^{x(1-t)} b\, b^{-1} \rp)^{1/(1-t) \mymod p} =
a^x,
\ee

\noindent
where $1/(1-t)$ can be computed modulo $p$ because we assumed $1<t<p$.

The case involving many source qudits, all of which can be used to conjugate
the target, is very similar to the above. The expression can be simplified
by moving all the $b$'s to the left, and combining similar factors. In the end,
the net effect will again be a series of $X$ and controlled-$X$ gates.

Finally, one may wonder about using an ancilla as an intermediate step.
That is, first we take an ancilla (say, $g'$), conjugate it by some function
(say, $f$) of some qudits, and then conjugate the target by the ancilla. 
However, the same effect can be achieved by conjugating the target first by
$f^{-1}$, then by $g'$ and finally by $f$. This procedure therefore provides
no extra computational power.

The conclusion is that the operations achievable from braiding magnetic
charges are exactly those generated by the $X$ and controlled-$X$ gates. 
In fact, the $X$ gate is redundant as we have assumed the existence of 
$\ket{1}$ ancillas, which can be used as control qudits in a 
controlled-$X$.

\subsection{\label{sec:fuse}Operations involving fusion of fluxes}

Now we turn to the operations achieved by the fusion of magnetic fluxes.
For these operations it will be sufficient to determine whether the two 
particles fused into the vacuum or not, thereby obtaining at most
one bit of information from each fusion.

At this point we remind the reader that standard states consist of pairs
of anyons, whose total flux is trivial. That is, the state $\ket{g}$ describes
an anyon of flux $g$ paired with an anyon of flux $g^{-1}$. There are 
therefore two basic choices for fusion: we can fuse the two anyons that compose
a single pair with each other, or we can fuse one of them with an anyon
from another pair, typically an ancilla. To avoid confusion, in the latter 
case we will always use the anyon of flux $g$ (rather than $g^{-1}$) 
for the fusion.

The case of fusion with an ancilla will lead to a measurement in the
$Z$ basis. The fusion of anyons from the same pair will lead to a measurement
in the $X$ basis. However, we will delay the construction of the actual
measurement gates until the next section. For this section, we will simply
describe the fusions as abstract operations on the computational
space by employing the construction of probabilistic projections.

The fusion of an anyon from a state $\ket{\Psi}$ with an anyon ancilla
of flux $b^{-1}$ is a probabilistic projection onto the subspace 
$\mathcal{K}=\{\ket{0}\}$. That is, an anyon of flux $a^i b a^{-i}$ can only
fuse into the vacuum with a flux  $b^{-1}$ if $i=0$ (modulo $p$ as usual).
When $i>0$ there must be an anyon left over to carry the non-trivial total 
flux. When $i=0$ the fusion can either produce the vacuum state or an anyon
with non-trivial charge. The probability for fusion into the vacuum in this 
case is $1/p$. Furthermore, if we fuse into the vacuum we can replace the state
with a $\ket{0}$ ancilla. Therefore the whole operation is a probabilistic 
projection onto $\ket{0}$ with $p_{PP}=1/p$.

The fusion of two anyons from the same pair is a probabilistic projection
onto the subspace $\mathcal{K}=\{\ket{\tilde 0}\}$. Because the total magnetic 
flux of the pair is always trivial, the fusion product must be an electric 
charge. The charge corresponds to a representation of $G$ given by the action 
of conjugation on the anyon fluxes. The state $\ket{\tilde 0}$ transforms 
trivially and corresponds to the vacuum, whereas the states $\ket{\tilde i}$,
for $i>0$, are orthogonal to the vacuum and correspond to non-trivial 
representations. In fact, this procedure is a probabilistic projection with
$p_{PP} = 1$. However, since the state is destroyed during fusion, to complete 
the projection we must be able to produce $\ket{\tilde 0}$ states. This will be
discussed below.

The other choices for fusion are equivalent to a combination of one of the
above measurements and an $X$ or controlled-$X$ gate.
Fusing with a flux of the form $a^{i}b^{-1} a^{-i}$ is equivalent to first 
applying a $X^{-i}$ gate and then performing a fusion with $b^{-1}$. 
A fusion with any other flux can never produce the vacuum if the qudit is in 
the computational subspace. 
Finally, one can consider fusion of anyons from two different qudits.
If the state of the two qudits is $\ket{i}\otimes\ket{j}$, the fusion
will only produce the vacuum state if $i=j$. Therefore, the operation can
be simulated by a controlled-$X^{-1}$, followed by the fusion of the target
with a $b^{-1}$ flux.

The conclusion so far is that fusion of magnetic charges provides us with
two new operations: the probabilistic projections onto the subspaces
$\ket{0}$ and $\ket{\tilde 0}$, which will eventually become measurements
in the $Z$ and $X$ bases. The only operation that has not been considered
is using the products of fusion for further operations or fusions. This subject
will be briefly touched upon after discussing fusion of electric charges.

\subsubsection{Production of $\ket{\tilde 0}$ states}

To conclude the discussion on fusion of fluxes, we present the construction
of $\ket{\tilde 0}$ states, which were needed to complete the probabilistic
projection onto $\ket{\tilde 0}$.

Just as the state $\ket{\tilde 0}$ naturally fuses into the vacuum, it
is also naturally produced from the vacuum. Unfortunately, producing a 
pair of anyons from the vacuum is just as likely 
to produce the vacuum state for one of the other superselection 
sectors as it is to produce the state $\ket{\tilde 0}$.
Therefore, after producing a vacuum state we must measure its 
superselection sector. Vacuum pairs that are produced in the computation 
subspace (magnetic charge in the conjugacy class of $b$) will be kept 
as $\ket{\tilde 0}$ states, and the rest will be tossed out.

Since measurements are done by fusion, which is a destructive procedure,
we must copy the vacuum state before measuring the conjugacy 
class. 
The procedure starts with a pair created from the vacuum and a $\ket{0}$ 
ancilla:
\be
\ket{\mbox{Vac}}\otimes\ket{0},
\ee

\noindent
and applies to it a swap, made out of the conjugation-based controlled-$X$:

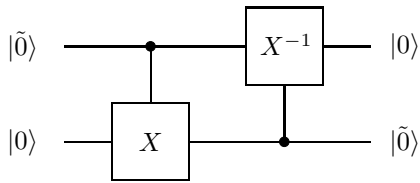
\begin{center}
\setlength{\unitlength}{1in}
\begin{picture}(3,1.1)(0,0)
{\put(.7,.8){\line( 1, 0){.95}}}
{\put(2.3,.8){\line( -1, 0){.25}}}
{\put(.7,.3){\line( 1, 0){.25}}}
{\put(2.3,.3){\line( -1, 0){.95}}}
{\put(1.15,.8){\line( 0,-1){.3}}}
{\thinlines \put(1.15,.8){\circle*{.05}}}
{\put(0.95,.1){\framebox(.4,.4){}}}
{\put(1.85,.3){\line( 0,1){.3}}}
{\thinlines \put(1.85,.3){\circle*{.05}}}
{\put(1.65,.6){\framebox(.4,.4){}}}
\put(1.15,.3){\makebox(0,0){$X$}}
\put(1.86,.81){\makebox(0,0){$X^{-1}$}}
\put(.4,.8){\makebox(0,0)[l]{$\ket{\tilde 0}$}}
\put(.4,.3){\makebox(0,0)[l]{$\ket{0}$}}
\put(2.4,.8){\makebox(0,0)[l]{$\ket{0}$}}
\put(2.4,.3){\makebox(0,0)[l]{$\ket{\tilde 0}$}}
\end{picture}
\end{center}

\noindent
where the circuit depicts the result for the case when the vacuum
pair was created in the computational superselection sector, in which case 
$\ket{\mbox{Vac}}=\ket{\tilde 0}$.

In the case when the vacuum state was not created in the computational
superselection sector, then the effect of the conjugations will be different. 
However, since the conjugations are performed using braiding, which never 
changes the superselection sector, the vacuum 
state can only be transformed into a state that is orthogonal to 
$\ket{0}=\ket{b}$.

After applying the above controlled-$X$ gates, we attempt to 
fuse an anyon from what was the vacuum state with an ancilla of flux $b^{-1}$.
If they fuse into the vacuum, this implies that the vacuum state was created
in the computational superselection sector, and the above circuit worked 
correctly. The ancillas $\ket{0}$ will have been transformed properly into a 
$\ket{\tilde 0}$ ancilla, which can be used for computation. In the case
when the fusion does not produce a vacuum state, the swap probably
did not produced the desired state, so we discard it and start over.

To summarize, we now have a source of $\ket{\tilde 0}$ ancillas, which can
be used as the last step needed to complete the probabilistic projection onto 
$\ket{\tilde 0}$.

\subsection{Representations and fusion of electric charges}

Thus far, we have only considered operations involving magnetic fluxes.
These operations led to a controlled-$X$, and measurements in the $X$ and
$Z$ bases. However, these gates do not form a universal gate-set. We must
therefore consider operations involving electric charges as well.

The electric charges transform as irreducible representation of the group $G$.
To obtain the spectrum of electric charges, as well as their braiding and
fusion rules, we must therefore discuss the representation theory of $G$.

It is easy to see that the commutator subgroup $G'$ of 
groups of the form $G=\Z_p \sdp{\theta} \Z_q$ is just $G'=\Z_p$.
The representation theory of $G$ can be obtained by inducing representations
from $G'$. Starting from the trivial representation on $G'$, the induced 
representations are the one-dimensional representations where $a\rightarrow 1$
and $b$ is a $q^{th}$ root of unity. 

The rest of the irreducible representations have dimension $q$
and are obtained by inducing from the non-trivial representations of $\Z_p$. 
The induced representations are all irreducible though not necessarily 
distinct. In fact, they can be easily described in their natural basis as 
\be
a \rightarrow \mypmatrix{\omega \cr & \omega^{t} \cr
             & & \ddots \cr  & & &\omega^{t^{q-1}}},
\ \ \ 
b \rightarrow \mypmatrix{0 & 1 \cr & 0  \cr 
             & & \ddots & 1 \cr  1 & & & 0},
\nonumber
\ee

\noindent
where $\omega$ is the $p^{th}$ root of unity of the representation from which
we are inducing. The matrix for $a$ is diagonal, whereas $b$ is the permutation
matrix with entries $1$ above the diagonal.

Even though the representation theory for these particular groups is easy,
we will use abstract language to describe the fusion rules, which will make
the connection to the general case clearer.

Take any non-abelian irreducible representation, and consider a pair of 
electric charges in the state $\ket{R(a^i)}_R$. 
What representations do we get if we fuse the 
two charges? The product of fusion is invariant under the action of $a$:
\be
U(a)\otimes U(a) \ket{R(a^i)}_R &=& \ket{R(a)R(a^i)R(a^{-1})}_R \nonumber\\
&=& \ket{R(a^i)}_R
\ee

\noindent
and therefore represents the commutator subgroup $G'$ by the identity.
This implies that the representation is abelian! In particular, it is
easy to see that the one-dimensional subspaces
\be
\ket{\lp[\gamma^j\rp]}_R \equiv 
\ket{\diag{\gamma^j,\gamma^{2j},\dots,\gamma^{q j}}}_R
\ee

\noindent
with $\gamma^q = 1$ are the spaces corresponding to the representations
$a\rightarrow 1$, $b \rightarrow \gamma^j$. 

We will be interested in the quantum amplitude that a state $\ket{R(a^i)}_R$ 
fuses into the $b \rightarrow \gamma^j$ representation. 
This quantity will be denoted by the fusion amplitude
\be
F_{i\rightarrow j} \equiv
\big\langle\lp[\gamma^j\rp]\big |R(a^i) \big\rangle_R
= \frac{1}{q} \sum_{k=1}^{q} \gamma^{-k j} \omega^{i t^{(k-1)}},
\ee

\noindent
with $0\leq i <p$ and $0\leq j <q$.

Let $\ket{\Psi}$ be an arbitrary state entangled with an electric charge pair
\be
\ket{\Psi} = \sum_{i=0}^{p-1} \ket{\Psi_i}\otimes\ket{R(a^i)}_R
\ee

\noindent
where the $\ket{\Psi_i}$ denotes (unnormalized) states of the rest of the
system. The fusion amplitudes allow $\ket{\Psi}$ to be rewritten as
\be
\ket{\Psi} = 
\sum_{i=0}^{p-1}\sum_{j=0}^{q-1}
 F_{i\rightarrow j}\ket{\Psi_i}\otimes\ket{\lp[\gamma^j\rp]}_{R} .
\ee

\noindent
The basis $\ket{[\gamma^j]}_R$ labels the total charge of the two
anyons that comprise the electric charge pair. A fusion of the two
electric charges, followed by a measurement of the resulting fusion product,
will be a measurement in this basis.

Note that the basis $\ket{[\gamma^j]}_R$ only spans the diagonal subspace
of $\ket{M}_R$. However, this is the subspace containing all the states
$\ket{R(a^i)}_R$. The subspaces spanned by $\ket{R(b^{j} a^i)}_R$, for some
fixed $j>0$, are mapped unchanged into the space of a single higher-dimensional
irreducible representation, and are therefore not useful for the purposes of 
this paper.

While the representation $R$ does not appear explicitly in the fusion 
coefficients, it enters implicitly in the above expression 
as the choice for $p^{th}$ root of unity $\omega$. Though we could use the 
notation $\omega_R$, this will not be necessary as we will generally 
work with only one higher-dimensional irreducible representation.

The most important feature of the $F_{i\rightarrow j}$ coefficients is 
that $\ket{R(a^{0})}_R$ is the vacuum state and therefore 
\be
F_{0\rightarrow j} = \delta_{j,0}
\ee

\noindent
which can be verified by direct calculation. Another important property is
that 
\be
\lp|F_{i\rightarrow j}\rp| > 0
\ee

\noindent
for all $i>0$. The proof involves showing that a linear relation of roots
of unity only vanishes if it is a combination of the obvious regular polygon
relations (which is proven in \cite{Shoenberg}).

A final interesting property is that
\be
F_{it^k\rightarrow j} = \gamma^{-j k} F_{i\rightarrow j},
\ee

\noindent
which is a consequence of
\be
\ket{R(a^{it^k})}_R &=& \ket{R(b^k a^i  b^{-k})}_R \nonumber\\
&=& U(b^k)\otimes U(b^k) \ket{R(a^i)}_R.
\ee

\subsection{Examples}

\subsubsection{$S_3$}

The group $S_3$ has three irreducible representations, the trivial (identity) 
representation (where $a\rightarrow 1$, $b\rightarrow 1$), 
the sign of the permutation 
(where $a\rightarrow 1$, $b\rightarrow -1$) and a two-dimensional one:
\be
a \rightarrow \mypmatrix{\omega & 0 \cr 0 & \bar \omega}, \ \ \ 
b \rightarrow \mypmatrix{0 & 1 \cr 1 & 0},
\ee

\noindent
where $\omega$ is a non-trivial cube root of unity.
The fusion amplitudes are
\be
F_{0\rightarrow0} &=& 1, \ \ \ 
F_{1\rightarrow0} = -\frac{1}{2}, \ \ \ 
F_{2\rightarrow0} = -\frac{1}{2}, \nonumber \\ 
F_{0\rightarrow1} &=& 0, \ \ \ 
F_{1\rightarrow1} = -i \frac{\sqrt{3}}{2}, \ \ \ 
F_{2\rightarrow1} = i \frac{\sqrt{3}}{2}.
\ee

The best way to visualize these coefficients is to start with a state 
$\ket{\tilde 0}$, and a pair of electric charges in the 
vacuum state of the two-dimensional representation: $\ket{R(I)}_R$. 
Then entangle with a controlled-sum to get
\be
\frac{1}{\sqrt{3}} \sum_j \ket{j} \ket{R(a^j)}_R
= \frac{1}{\sqrt{3}} \sum_j \ket{j} 
\bket{\mypmatrix{\omega^j & 0 \cr 0 & \bar \omega^j}}_R.
\ee

Fusion of the electric charge pair produces either the vacuum (trivial
representation) or a charge transforming under the sign representation. 
The probability of getting each is
\be
P_{vac} &=& \sum_j \lp|\frac{1}{\sqrt{3}} F_{j\rightarrow 0} \rp|^2 
= \frac{1}{2}, \nonumber\\
P_{sgn} &=& \sum_j \lp|\frac{1}{\sqrt{3}} F_{j\rightarrow 1} \rp|^2 
= \frac{1}{2},
\ee

\noindent
and the state of the magnetic charges afterwards is one of
\be
\ket{\Psi_{vac}} &=& \frac{1}{\sqrt{6}} 
\lp( 2 \ket{0} - \ket{1} - \ket{2} \rp), \nonumber\\
\ket{\Psi_{sgn}} &=& \frac{1}{\sqrt{2}} \lp( \ket{1} - \ket{2} \rp).
\ee

\noindent
These are obtained by multiplying the initial state by the appropriate $F$
coefficients and renormalizing to unit magnitude. In the case of the second
state we also introduced an extra global phase of $i$, which is related
to the arbitrary choice of phase of the $\ket{[\gamma^j]}_R$ states.

\subsubsection{$\Z_7 \sdp{\theta} \Z_3$}

The group $\Z_7 \sdp{\theta} \Z_3$ has five irreducible representations. 
Three of them
are one dimensional and set $a\rightarrow 1$ and $b$ to a cube root of unity.
The other two are three dimensional and are complex conjugates of each other.

The main new feature of this group is that the non-trivial powers of $a$
are not all conjugate to one another. This leads to more complicated 
fusion coefficients. For example:
\be
F_{0\rightarrow 1}&=& 0,
\nonumber\\*
F_{1\rightarrow 1}&=& \frac{1}{3} A,\ \ \ \ 
F_{2\rightarrow 1}= \frac{\gamma^{2}}{3} A,\ \ \ \ 
F_{3\rightarrow 1}= \frac{\gamma}{3} B,
\nonumber\\*
F_{4\rightarrow 1}&=& \frac{\gamma}{3} A,\ \ \ \ 
F_{5\rightarrow 1}= \frac{\gamma^{2}}{3} B,\ \ \ \ 
F_{6\rightarrow 1}= \frac{1}{3} B,
\ee

\noindent
with
\be
A &=& \gamma^{2}\omega + \gamma \omega^2 + \omega^4  
\nonumber\\*
  &=& \e{2\pi i \frac{17}{21}} + \e{2\pi i \frac{13}{21}} + 
      \e{2\pi i \frac{12}{21}},
\nonumber\\
B &=& \gamma^{2}\omega^{-1} + \gamma \omega^{-2} + \omega^{-4}
\nonumber\\*
  &=& \e{2\pi i \frac{11}{21}} + \e{2\pi i \frac{1}{21}} + 
      \e{2\pi i \frac{9}{21}},
\ee

\noindent
where we have chosen $\gamma=\e{2\pi i/3}$ and $\omega=\e{2\pi i/7}$.
Notice how $A$ is close in magnitude to $3$ whereas $B$ is close in magnitude 
to $1$.

\subsection{\label{sec:Zelec}Operations involving electric charges}

Now it is time to apply the discussion in the previous subsections to build
a useful operation out of electric charges.
While there seems to be a wealth of strange ancillas that could be produced
using electric charges, most of them have complicated relative
amplitudes or phases that are hard to use in a constructive proof of
universal computation. We will therefore focus our attention on producing
an operation that arises naturally from the fusion amplitudes: the projection 
onto the subspace orthogonal to $\ket{0}$.

Consider a qudit in the state
\be
\ket{\Psi} = \sum_{i=0}^{p-1} \psi_i \ket{i},
\ee

\noindent
where the coefficients $\{\psi_i\}$ could either be complex numbers, or 
could represent the state of the rest of the system if the qudit is 
entangled with other qudits.

We append to the qudit an electric charge pair $\ket{R(I)}_R$ in the vacuum
state of a non-abelian representation $R$. Using braiding, we can right 
multiply the state of the electric charge by some function $f$ of the qudits 
flux:
\be
\ket{\Psi}\otimes \ket{R(I)}_R \longrightarrow 
\sum_{i=0}^{p-1} \psi_i \ket{i} \otimes \ket{R(f(i))}_R.
\ee

We have shown in Section~\ref{sec:braid} 
that the most general function is of the form 
$f(i) = a^{\alpha} a^{\beta i} b^{\delta}$. Choosing $\delta\neq 0$ turns out
not to be useful, and choosing $\alpha\neq 0$ can be used to get projections
to the spaces orthogonal to $\ket{i}$ for $i>0$, but this can be achieved as 
well with an $X$ gate. We will therefore focus on $f(i) = a^{\beta i}$ so
that we obtain the state
\beq
\sum_{i=0}^{p-1} \psi_i \ket{i} \otimes \ket{R(a^{\beta i})}_R
= \sum_{i=0}^{p-1}\sum_{j=0}^{q-1}
F_{\beta i\rightarrow j} \psi_i \ket{i} \otimes \ket{\lp[\gamma^j\rp]}_R.
\eeq

A fusion of the electric charge pair, followed by a measurement of the 
resulting electric charge (the feasibility of which will be the subject of 
Section~\ref{sec:1de} below), leads to a state that is proportional to
\be
\sum_{i=0}^{p-1} F_{\beta i\rightarrow j} \psi_i \ket{i},
\ee

\noindent
where $j$ now labels the result of the measurement in the basis 
$\ket{\lp[\gamma^j\rp]}_R$.

Because of the property $F_{0\rightarrow j} = \delta_{j,0}$, if the measurement
result is $j\neq 0$, we will have projected into the space orthogonal
to $\ket{0}$. Unfortunately, we will have also introduced undesired
relative phases and amplitudes. The trick will be to balance these out.

Consider repeating the above procedure $p-1$ times, with $\beta$ taking
values from $1$ to $p-1$. Furthermore, assume that in each case
the fusion results in $j=1$. The resulting state will be, up to normalization:
\be
\propto  \sum_{i=1}^{p-1} \lp(\prod_{\beta=1}^{p-1} 
F_{\beta i\rightarrow 1} \rp) \psi_i \ket{i}
\propto \sum_{i=1}^{p-1} \psi_i \ket{i},
\ee

\noindent
where we have used the fact that multiplication by $i$, modulo $p$, is just
a rearrangement of the values of $\beta$. 

The above procedure is a probabilistic projection onto 
$\mathcal{K}=\ket{0}^{\perp}$. As usual, if we do not obtain $j=1$ as the 
result of each measurement, we just discard the state being projected.

What is the probability of success of the above procedure? The probability 
for obtaining $j=1$ on the first try is
\be
P_{j=1} = \sum_{i=1}^{p-1} \lp| F_{i\rightarrow 1} \psi_i \rp|^2
\geq \underset{i>0}{\text{min}} \lp( \lp| F_{ i\rightarrow 1} \rp|^2 \rp )
\sum_{i=1}^{p-1} \lp| \psi_i \rp|^2.
\ee

On subsequent measurements, the state has previously been projected 
to $\ket{0}^\perp$ and renormalized.
Therefore the probability of success for each trial is simply bounded by
\be
P_{j=1} 
\geq \underset{i>0}{\text{min}} \lp( \lp| F_{ i\rightarrow 1} \rp|^2 \rp ).
\ee

The total probability of success is just the product of these quantities.
In particular, the probability $p_{PP}$ associated with the 
probabilistic projection can be bounded by
\be
p_{PP} \geq 
\underset{i>0}{\text{min}} \lp( \lp| F_{ i\rightarrow 1} \rp|^2 \rp )^{p-1}
> 0,
\ee

\noindent
where we used the fact that $| F_{ i\rightarrow j}|>0$ for $i>0$.

Of course, the above is a underestimation of the probability of obtaining
a good projection. For example, if all the results $j$ were equal to some
fixed $j>1$, the same argument would show that a correct projection was 
obtained.
Furthermore, there are many other ways in which the relative phases and
amplitudes can cancel out. A classical computer, with knowledge of the
values of $F_{ i\rightarrow j}$, can keep repeating the procedure until such
a cancellation occurs. The computer would also be required to stop after a long
sequence of $j=0$ results, in which case the state would have been projected 
onto $\ket{0}$.

In the end, as long as $p_{PP}$ is fixed and finite, we have produced
the desired probabilistic projection to the space $\ket{0}^\perp$. Different
values of $p_{PP}$ will just affect the complexity of an algorithm as
a multiplicative constant. Furthermore, for the small groups that are likely
to appear in the laboratory, $p_{PP}$ should be reasonably large. For example,
in the case of $G=S_3$, $p_{PP}$ can be made exponentially close to one in 
the number of measurements.

It should be noted that because we are working with qudits of dimension
$d=p$, and
the semidirect product requires $p>q\geq 2$, the above projection will always
be a non-trivial operation. In fact, it will always be powerful enough to
complete a universal gate-set.

At this point, all that remains to be done is to prove the universality of 
the gates constructed from the basic anyon operations. This will be the subject
of Section~\ref{sec:gates}. However, before closing this section, we shall
discuss some issues regarding the measurability of electric charges, and look
at some alternative operations that could have been employed.

\subsection{\label{sec:1de}On the measurement of 
one-dimensional representations}

The feasibility and accuracy of the probabilistic projection onto 
$\ket{0}^\perp$ depend crucially on
being able to identify electric charges carrying 
one-dimensional representations. However, these charges have a special 
property that makes them hard to identify: when only using braiding,
a one-dimensional representation is indistinguishable from the vacuum!

The reason behind the above difficulty is that one-dimensional representations
of a group $G$ are constant on conjugacy classes of $G$. Therefore,
a magnetic charge that is braided around one of these electric charges will
have its state change by an overall phase. These global phases are
not measurable in quantum mechanics.

Of course, an interference experiment would produce a measurement of the
charge. The standard double-slit experiment, with the electric charge located
in between the slits, will produce a pattern on the screen that depends 
on the representation of the electric charge. However, during the experiment, 
the anyon will be in a superposition of spatial positions which is no longer 
protected from decoherence by topology. Since the interference experiment
can be repeated many times without affecting the electric charge, 
this may not necessarily be a problem. However, it does involve working in
a regime where the anyons can be treated as waves rather than particles.

On the bright side, these electric charges can also be detected by fusion,
assuming the availability of electric charge ancillas with one-dimensional 
representations. Their fusion rules are particularly simple because 
these states have a one-dimensional internal Hilbert space. Furthermore, 
their fusions always produce unique results. If $\gamma(g)$ and
$\gamma'(g)$ are two one-dimensional representations of a group $G$,
then the fusion of the electric charges carrying these representations 
produces a charge of representation $\gamma''(g)=\gamma(g)\gamma'(g)$.
A charge will only fuse into the vacuum when fused with its conjugate 
representation. Therefore, after a series of fusions that end up producing
the vacuum state, we can determine the representation of the original
electric charge.

In fact, for groups with $q=2$ such as $S_3$, there is a further 
simplification.
In these groups there are only two one-dimensional representations: the vacuum
and the sign representation. Since the fusion of $\ket{R(a^i)}_R$ produces 
a one-dimensional charge, if it does not fuse into the vacuum, then it must
have produced the sign charge. Therefore, for these groups, we do not even
require one-dimensional electric charge ancillas.

\subsection{\label{sec:alt}Other possibilities}

In this section, we will briefly discuss one last possibility
for producing useful operations: using the products of fusion. Though not
strictly needed to complete a universal gate-set, this subsection is an 
interesting study of alternative operations and the effects of decoherence
during fusion.

At first sight, it appears that the projection onto $\ket{0}^\perp$ 
can be done without using electric charges with the following procedure:
first fuse one anyon from the state to be measured with a $b^{-1}$ flux. 
Only the $\ket{0}$ state can fuse into the vacuum. If an anyon remains, 
fuse again with a $b$ flux to restore it to its previous state, and pair 
it with its old partner. Repeating
the procedure multiple times (because the $\ket{0}$ could turn into an
electric charge rather than the vacuum) yields the desired projection.

There are, however, two problems with the above construction. The first, and
smaller, problem is that when fusing with $b^{-1}$ or with $b$ we could be
turning our magnetic charges into dyons. For groups of the form 
$\Z_p \sdp{\theta} \Z_q$ the dyonic electric charges are all one dimensional,
however, and will therefore have no effect on braiding, as discussed in the
previous section. The probabilities of fusion into the vacuum will be reduced,
and therefore so will the respective projection probabilities,
but they will still remain non-zero. In fact, a careful examination of the
operations constructed so far shows that they work with a probabilistic 
mixture of dyons and regular magnetic charges.

The second and larger problem, though, is decoherence. The fluxes 
$a^i b a^{-i}b^{-1}= a^{i(1-t)}$ belong, in general, to different conjugacy
classes and therefore different superselection sectors. When the quantum 
state is encoded in this form, it is susceptible to decohere into the
different superselection sectors.

When does this decoherence occur? It occurs during fusion. In general, fusion
takes two $n$-dimensional Hilbert spaces $\mathcal{H}$ and maps them to one:
$\mathcal{H}_1\times\mathcal{H}_2\rightarrow\mathcal{H}_3$. But quantum 
mechanics is unitary; therefore, what must really be happening is a mapping
to a tensor product of $\mathcal{H}$ and the environment:
$\mathcal{H}_1\times\mathcal{H}_2\rightarrow\mathcal{H}_3\times\mathcal{E}$.
When two states are mapped onto new states that are orthogonal in the
environment subspace, decoherence occurs.

How do we know if states will have orthogonal environment components after 
fusion? If two states belong to the same superselection sectors, they are
related by symmetry, which protects them from decoherence. This may not be
the case when they come from different superselection sectors, though.

For example, consider the states $\ket{a^i}\otimes\ket{a^j b}$ for $i$ and $j$
between $0$ and $p-1$, where the kets will denote single anyons in this 
paragraph and the next. States of different $j$ are all in the same conjugacy 
class, but states of different $i$ are grouped into conjugacy classes of $q$
elements (except for $i=0$, which is its own conjugacy class). In total, we
are talking about $p^2$ states.

These states fuse into the states with flux $a^k b$ for $0\leq k<p$. The 
resulting states may also have one of $q$ electric charges. In total, we fuse
into a space containing $p q$ states. Since $p q < p^2$, what must be happening
is that different conjugacy classes are mapped to states that are orthogonal 
in the environment subspace.

Note that the decoherence seems to occur when fusing out of a state made
up of different superselection sectors. However, fusion is the only 
operation that could have measured the relative phase between the sectors,
and it clearly does not. Therefore, it is acceptable to assume that the
decoherence occurs as soon as states are mapped into different superselection 
sectors.

Returning to the question of alternative implementations of the projection
onto $\ket{0}^\perp$, it is clear that the procedure described above
does not achieve its goals without causing decoherence in the general case.
However, in the special case when $q=p-1$, the non-trivial powers of $a$ 
form one conjugacy class. Therefore, the above trick can produce a 
projection onto $\ket{0}^\perp$ using only magnetic charges. Of course,
$q=p-1$ only holds for $G=S_3$.

For other groups, the operation could
become useful if we could tell into which superselection sector the state 
decohered, producing a probabilistic projection onto a smaller space.
The smaller projections may also be computationally powerful. However,
since we have completed a universal gate-set without the results of this 
subsection, we shall work on proving universality from the previously 
constructed gates, rather than pursuing this matter further.

\section{\label{sec:gates}Gate-set Universality}

The goal of this section is to prove the universality of the following 
qudit gate-set, which includes measurements:

\begin{enumerate}
\item Controlled-$X$,
\item Probabilistic projection onto $\ket{0}$,
\item Probabilistic projection onto $\ket{\tilde 0}$,
\item Probabilistic projection onto $\ket{0}^\perp$,
\end{enumerate}

\noindent
where we assume that the qudits are of dimension $d>2$, with $d$ prime.
The first requirement on $d$ is needed to make the gate-set universal,
whereas the second one will allow us to relate this gate-set to Gottesman's
gate-set \cite{Gottesman:1998se}. The above gate-set must be 
supplemented by a controlling computer capable of universal classical 
computation.

The above gates were selected as those arising naturally from the anyons
based on the groups $\Z_p \sdp{\theta} \Z_q$. The proof of universality 
of the above gate-set is the last step needed to show that universal
quantum computation is feasible with these anyons.

The proof of universality will proceed in two steps. In the first step
we will turn the second and third gates into proper measurements in the 
$Z$ and $X$ bases.  Most of the methods of the first step were described 
while building computation with non-solvable anyons \cite{me2002}. 
The second step involves using the probabilistic projection onto 
$\ket{0}^\perp$ to construct magic states that complete the universal 
gate-set. This is the new element needed to achieve universality 
with solvable anyons.

\subsection{\label{sec:gates0}Non-destructive measurement of $Z$ and $X$}

By the end of this subsection we will have constructed measurements in
the $Z$ and $X$ bases. These measurements will be non-destructive in the
sense that if result $i$ was obtained, the measured qudit will be in state
$\ket{i}$ or $\ket{\tilde i}$ respectively. 
Because the measurements in question are complete, the
non-destructive requirement can be achieved by having ancillas
for every eigenstate of $X$ and $Z$, and then using the controlled-$X$
to swap the ancillas into the computational space.

The construction begins by producing a set of basic ancillas. Along the way
we will also produce the $X$ and $Z$ unitary gates.

\subsubsection{$\ket{0}$ and $\ket{\tilde 0}$ ancillas}

Clearly, given $\ket{0}$ ancillas we can use the third gate to produce 
$\ket{\tilde 0}$ ancillas. 
Similarly, given $\ket{\tilde 0}$ ancillas we can use the second gate to
produce $\ket{0}$ ancillas. Therefore, if the initial state of the
quantum computer overlaps with either state, we can produce both kinds of 
ancilla.

Usually, the initial state of the quantum computer is $\ket{0}$. However,
by using the controlled-$X$ gate, in combination with the projections onto
$\ket{0}$, we can obtain these states no matter what
the qudits are initialized to. The procedure
is just to apply a controlled-$X^{-1}$ (equivalent to $d-1$ controlled-$X$ 
gates) to two qudits, and then project the target to the $\ket{0}$ space.
If the initial state had some overlap with any of the states 
$\ket{i}\otimes\ket{i}$, then this produces the desired ancillas. 
Furthermore, even if we allow 
states that are initially entangled, once we involve more than $d$ qudits, 
at least one pair must have an overlap with the diagonal states. Therefore,
$\ket{0}$ states can always be produced.

Henceforth, we shall assume an ample supply of $\ket{0}$ and $\ket{\tilde 0}$ 
ancillas.

\subsubsection{$\ket{1}$ states, $\ket{\tilde 1}$ states; $X$ gates, $Z$ gates}

The next step is to produce $\ket{1}$ and $\ket{\tilde 1}$ ancillas. The 
importance of these ancillas is that they will break the symmetry currently
present in the one-qudit Hilbert space. 

There are two symmetries in the Hilbert space that are not fixed by the
basic four gates of our set. The first symmetry is a relabeling
$\ket{ix}\rightarrow\ket{i}$, calculated modulo $d$, for some $0<x<d$. 
The second, is the relabeling $Z^{y}\rightarrow Z$,
for integer $0<y<d$. For fixed $x$, the second symmetry is a relabeling of our 
$d^{th}$ root of unity $\omega$ by $\omega^y \rightarrow \omega$ and a 
relabeling $|\widetilde{j y}\rangle\rightarrow\ket{\tilde{j}}$.

Therefore, given an ancilla in a state $\ket{x}$, with $x>0$, we can just 
rename it so that it becomes a $\ket{1}$ ancilla.
Similarly, given an ancilla in a state $\ket{\tilde y}$, $y>0$, 
we can relabel it as $\ket{\tilde 1}$. In fact, both can be done 
simultaneously in a consistent fashion, even if we do not know the
values of $x$ and $y$.

The initial states $\ket{x}$ and $\ket{\tilde y}$ can be obtained
from two maximally mixed states. The maximally mixed states can be described 
either as a state $\ket{x}$ with $x$ chosen at random, or a state 
$\ket{\tilde y}$ with $y$ chosen at random. Therefore, two maximally mixed
states serve our purpose as long as we do not obtain $x=0$ or $y=0$. These
two bad cases will be detected below, in which case the process can be 
restarted with two new mixed states.

To produce the maximally mixed states we apply a controlled-$X$ with 
$\ket{\tilde 0}$ as source and $\ket{0}$ as target.
The result is a maximally entangled state, which can be turned into a 
maximally mixed state by discarding one of the two qudits. Two of these
mixed states will serve as our ancillas.

Given our two ancillas, which we have now labeled $\ket{1}$ and 
$\ket{\tilde 1}$, we can build $X$ and $Z$ gates which are consistent
with the new labeling.
The $X$ gate is clearly just a controlled-$X$ with a $\ket{1}$ state as 
control, whereas the $Z$ gate is just a controlled-$X$ with a 
$\ket{\tilde 1}$ as target. The less familiar second construction is just a
specific case of the following circuit:

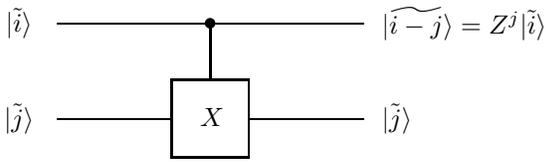
\begin{center}
\setlength{\unitlength}{1in}
\begin{picture}(3.2,1)(0.2,0)
{\put(.7,.8){\line( 1, 0){1.6}}}
{\put(.7,.3){\line( 1, 0){.6}}}
{\put(2.3,.3){\line( -1, 0){.6}}}
{\put(1.5,.8){\line( 0,-1){.3}}}
{\thinlines \put(1.5,.8){\circle*{.05}}}
{\put(1.3,.1){\framebox(.4,.4){}}}
\put(1.51,.31){\makebox(0,0){$X$}}
\put(.5,.8){\makebox(0,0){$\ket{\tilde i}$}}
\put(.5,.3){\makebox(0,0){$\ket{\tilde j}$}}
\put(2.4,.8){\makebox(0,0)[l]{$|\widetilde{i-j}\rangle 
= Z^{j} \ket{\tilde i}$}}
\put(2.4,.3){\makebox(0,0)[l]{$\ket{\tilde j}$}}
\end{picture}
\end{center}

At this point, if we were unlucky enough to get $x=0$ or $y=0$, then
one of the transformations $X$ or $Z$ will be the identity operator. This
can easily be checked by applying them to $\ket{0}$ or $\ket{\tilde 0}$ 
ancillas and then using the available probabilistic projections.

The $X$ and $Z$ gates can also be used to produce a reservoir of $\ket{1}$ 
and $\ket{\tilde 1}$ ancillas that will be consistent with the original
states. Two elements in the reservoir can also be compared, for example, by 
applying a $Z$ built from one ancilla followed by a $Z^{-1}$ built from the 
other. Therefore, even if the states were to decay over time, by using 
majority voting the damaged states can be weeded out.

In some cases, the one-qubit Hilbert spaces do have natural $\ket{1}$
or $\ket{\tilde 1}$ states, which implies a natural way of measuring or
obtaining such states. For those systems, either the natural ancillas or
the arbitrary ones constructed above can be used. For example, for
the anyons $\ket{1}=\ket{a b a^{-1}}$. However, choosing a different
$\ket{1}$ state is equivalent to choosing a different element $a$.

\subsubsection{Measurements of $Z$ and $X$}

At this point all the elements are in place to produce measurements
in either the $Z$ basis or the $X$ basis.

The key element of the $X$ basis measurement is the circuit

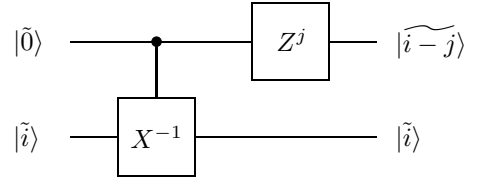
\begin{center}
\setlength{\unitlength}{1in}
\begin{picture}(3,1.1)(0,0)
{\put(.7,.8){\line( 1, 0){.95}}}
{\put(2.3,.8){\line( -1, 0){.25}}}
{\put(.7,.3){\line( 1, 0){.25}}}
{\put(2.3,.3){\line( -1, 0){.95}}}
{\put(1.15,.8){\line( 0,-1){.3}}}
{\thinlines \put(1.15,.8){\circle*{.05}}}
{\put(0.95,.1){\framebox(.4,.4){}}}
{\put(1.65,.6){\framebox(.4,.4){}}}
\put(1.15,.3){\makebox(0,0){$X^{-1}$}}
\put(1.86,.81){\makebox(0,0){$Z^j$}}
\put(.4,.8){\makebox(0,0)[l]{$\ket{\tilde 0}$}}
\put(.4,.3){\makebox(0,0)[l]{$\ket{\tilde i}$}}
\put(2.4,.8){\makebox(0,0)[l]{$|\widetilde{i-j}\rangle$}}
\put(2.4,.3){\makebox(0,0)[l]{$\ket{\tilde i}$}}
\end{picture}
\end{center}

\noindent
applied to a $\ket{\tilde 0}$ ancilla and the state to be measured. 
If the above circuit is repeated many times, each time with a different 
$\ket{\tilde 0}$ ancilla, and with $j$ varying from $0$ to $d-1$,
we obtain the transformation
\be
\sum_i \beta_i \ket{\tilde i} \longrightarrow \sum_i \!\!&\beta_i&\!\! 
\ket{\tilde i} \otimes \ket{\tilde i}
\cdots \otimes |\widetilde{i-1}\rangle 
\otimes |\widetilde{i-1}\rangle \cdots 
\nonumber\\*
&&\otimes |\widetilde{i-d+1}\rangle.
\ee

A probabilistic projection onto $\ket{\tilde 0}$ can then be applied to
each qudit. If one of the qudits of the form $|\widetilde{i-j}\rangle$
projects onto the space $\ket{\tilde 0}$, then the outcome of the measurement
is $j$.

Note that because of the one-sided error model of the probabilistic projection,
an erroneous measurement result can never be obtained, no matter how small 
$p_{PP}$ is. The worst possible outcome is that after all the qudits have 
been measured, no conclusion can be reached. Of course, a standard small
two-sided probability of error can also be made exponentially small by using
enough qudits in the above measurement. 

The measurement in the $Z$ basis proceeds similarly, where the transformation
\be
\sum_i \alpha_i \ket{i} \longrightarrow \sum_i \!\!&\alpha_i&\!\!
\ket{i} \otimes \ket{i}
\cdots \otimes \ket{i-1} \otimes \ket{i-1} \cdots
\nonumber\\*
&&\otimes \ket{i-d+1}
\ee

\noindent
is performed using the $X$ and controlled-$X$ gates, followed by a 
probabilistic projection onto $\ket{0}$.

Finally, the above measurements can be performed non-destructively, by 
projecting all but one of the qudits. 
Alternatively, the eigenstates of $X$ and $Z$ can be directly constructed from 
these gates and $\ket{0}$ or $\ket{\tilde 0}$ eigenstates.

\subsection{\label{sec:magic}Completing the gate-set}

So far, we have only shown that our gates can realize operations in the 
Clifford group. In order
to achieve universal quantum computation we need to complete the gate-set
with an operation outside the Clifford group. 

It was shown in Ref.~\cite{me2002} that the Toffoli gate, combined with 
measurements in the $X$ and $Z$ bases, is universal for quantum computation.
Therefore, a successful construction of the Toffoli out of our gate-set
will prove it universal. The Toffoli gate will be constructed out of the
previously described operations, together with the thus far unused
probabilistic projection onto $\ket{0}^\perp$.

In addition to producing measurement gates, probabilistic projections are 
particularly useful for preparing magic states, which are ancillas whose 
use allows us to perform new gates such as the Toffoli. In particular, we 
shall show that we can produce the two magic states
\be
\ket{\phi_{M1}} &=& \frac{1}{d}
\sum_{i,j} \ket{i}\otimes\ket{j}\otimes\ket{i j},\nonumber\\
\ket{\phi_{M2}} &=& \frac{1}{d}
\sum_{i,j} \omega^{\delta_{i,0}\delta_{j,0}} \ket{i}\otimes\ket{j},
\ee

\noindent
where $\delta_{i,j}$ is the Kronecker delta function.
The first of these states produces the Toffoli gate up to some errors in the
Clifford group. The second magic state allows us to correct these errors,
and in fact, allows the construction of the complete Clifford group
even without the use of the first magic state. 

We shall begin by discussing how to use each of the magic states, and then 
afterwards turn to the task of describing their construction out of the
available operations.

\subsubsection{Using $\ket{\phi_{M1}}$}

The magic state $\ket{\phi_{M1}}$ and its use in producing the Toffoli gate
was first introduced by Shor \cite{Shor:1996qc}, and generalized to qudits
in Ref.~\cite{Gottesman:1998se}. We shall give a brief description of its use
in order to give an account of the exact Clifford group operations needed
in the last step as corrections.

The procedure begins with a general state
\be
\ket{\Psi}=\sum_{a,b,c} \psi_{a,b,c} \ket{a}\otimes\ket{b}\otimes\ket{c},
\ee

\noindent
to which an ancilla $\ket{\phi_{M1}}$ is appended. A controlled-$X^{-1}$
is applied to the first data qudit with the first ancilla qudit as control.
Similarly, a controlled-$X^{-1}$ is applied to the second data qudit
from the second ancilla qudit, and a controlled-$X$ is applied to the third
ancilla qudit, from the third data qudit. The first two data qudits are then 
measured in the $Z$ basis and the third data qudit is measured in the $X$
basis. If the results of the measurements are $\alpha$, $\beta$ and $\gamma$
respectively, then the remaining qudits are left in the state
\beq
\sum_{a,b,c} \psi_{a,b,c}\, \omega^{\gamma c}\,
\ket{a-\alpha}\otimes\ket{b-\beta}
\otimes\ket{(a-\alpha)(b-\beta)+c}.
\eeq

The corrections begin by applying an 
$X^\alpha\otimes X^\beta\otimes X^{-\alpha\beta}$ gate followed by 
a controlled-$X^\beta$ from the first qudit to the third qudit and
a controlled-$X^\alpha$ from the second qudit to the third qudit. The
state then becomes
\be
\sum_{a,b,c}\psi_{a,b,c}\, \omega^{\gamma c}\,
\ket{a}\otimes\ket{b}
\otimes\ket{a b + c}.
\ee

All that is needed to complete the Toffoli gate is a $Z^{-\gamma}$ gate 
applied to the third qudit and a phase $\omega^{\gamma a b}$ applied to 
the first two qudits.
Unfortunately, we must first build the latter transformation out of
the second magic state.

\subsubsection{Using $\ket{\phi_{M2}}$}

Once again, the magic state is appended to a pair of qudits. Now 
controlled-$X$ gates are applied with the data qudits as source and the 
ancilla qudits as targets. 
Then the ancilla qudits are measured in the computational basis. The
outcomes $\alpha$ and $\beta$ will be uniformly distributed, and at 
the end we will have produced the transformation
\be
\sum_{a,b}\psi_{a,b} \ket{a}\otimes\ket{b}
\longrightarrow
\sum_{a,b}\psi_{a,b}\, \omega^{\delta_{a,\alpha}\delta_{b,\beta}}\, 
\ket{a}\otimes\ket{b}.
\ee

This procedure randomly and uniformly chooses a computational basis 
state and multiplies it by a phase of $\omega$. Repeated application of this 
transformation will eventually yield any of the $d^{d^2}$ states of the form
\be
\sum_{a,b}\psi_{a,b}\, \omega^{f(a,b)}\, 
\ket{a}\otimes\ket{b},
\ee

\noindent
where $f$ is an arbitrary  integer-valued function. 
This process is effectively a classical
random walk on a $d^2$ dimensional periodic lattice with $d^{d^2}$ nodes, 
where each use of a magic state is equivalent to taking one step. Because
the lattice is finite, after a polynomially large number of steps the
probability of not having arrived at least once at any one of the above states
becomes exponentially small. 

The final correction needed to complete the Toffoli gate was the phase
transformation to the state with $f(a,b)=\gamma a b$, and
can therefore be realized using many copies of the second magic state. 
All that remains
to prove universality is to describe the production of the magic states.

\subsubsection{Making the magic states}

The final piece of the puzzle is the production of the magic states using
the probabilistic projection onto $\ket{0}^\perp$. 

Probabilistic projections onto a subspace are particularly powerful for making
magic states, because it can be assumed that they successfully project into
the subspace every time. That is, if the probabilistic projection does not 
project onto the desired subspace, the state is tossed out and the procedure
is restarted from the beginning. Therefore, the probabilistic projection onto 
$\ket{0}^\perp$ effectively takes a state and removes the $\ket{0}$ component
of the state:
\be
\sum_{i=0}^{d-1} \alpha_i \ket{i} \longrightarrow 
A \sum_{i=1}^{d-1} \alpha_i \ket{i},
\ee

\noindent
where $A$ is some normalization constant. In fact, by combining this projection
with the $X$ gate, we can remove any of the components $\ket{i}$.

The main strategy for this section is to construct a series of ancilla states
of increasing complexity, until finally the desired magic states are obtained.
At this point, we have a supply of ancillas of the form $\ket{i}$ and 
$\ket{\tilde j}$ for any $i$ and $j$. From the $\ket{\tilde 0}$ state we can 
also make the ancilla $(\ket{0} + \ket{1})/\sqrt{2}$ by removing all 
$\ket{i}$ for $i>1$ with the probabilistic projection.

The next step is to produce entangled two-qudit ancillas. Given a supply
of ancillas of the form $\ket{\Psi} = \sum_i \psi_i \ket{i}$ we shall
produce ancillas of the form
\be
\ket{\Psi'} &=& 
\lp( \psi_0 \ket{0}\otimes\ket{1} + 
\sum_{i=1}^{d-1} \psi_i \ket{i}\otimes\ket{0} \rp) \nonumber\\
&=& \sum_{i=0}^{d-1} \psi_i \ket{i}\otimes\ket{\delta_{i,0}}.
\ee

The procedure begins with the state
\beq
\ket{\Psi}\otimes 
\lp(\frac{1}{\sqrt{2}} \ket{0} + \frac{1}{\sqrt{2}} \ket{1} \rp)
= \frac{1}{\sqrt{2}} 
\sum_{i=0}^{d-1} \sum_{j=0}^{1} \psi_i \ket{i}\otimes\ket{j},
\eeq

\noindent
which in general has $2d$ non-zero coefficients. We need to remove $d$
of these coefficients to obtain the state $\ket{\Psi'}$.

The procedure, done once for each $k$ from $1$ to $d-1$, is the 
following: First, apply a
controlled-$X^k$ with the left qudit as source and the right qudit as target.
Then, the right qudit is projected onto $\ket{0}^\perp$, and finally
the controlled-$X^k$ is undone. 
For each $k$, we remove the components $\ket{0}\otimes\ket{0}$ and 
$\ket{-1/k}\otimes\ket{1}$. The operation $-1/k$ is modulo $d$ as usual,
and ranges over all integers between $1$ and $d-1$ because $d$ is prime. 
Therefore, given a supply of $\ket{\Psi}$ ancillas, we can probabilistically
convert some of them into a supply of $\ket{\Psi'}$ ancillas.

Note that the above procedure works even if the coefficients $\psi_i$
represent the state of other qudits, as long as these are ancilla qudits
that can be tossed out if the projection procedure fails. In the same spirit,
given ancillas of the form
\be
\ket{\Phi} = \sum_{i=0}^{1} \sum_{j=0}^{1} \phi_{i,j} \ket{i}\otimes\ket{j},
\ee

\noindent
we can produce the three-qudit ancillas
\be
\ket{\Phi'}= \sum_{i=0}^{1} \sum_{j=0}^{1} \phi_{i,j} \ket{i}\otimes\ket{j}
\otimes\ket{\delta_{i,0}\delta_{j,0}}.
\ee

The procedure again involves appending $(\ket{0} + \ket{1})/\sqrt{2}$
to the ancilla $\ket{\Phi}$, which now generically has eight non-zero 
coefficients, and removing four of them. This is done with a set of 
controlled-$X$ gates with the third qudit as target, followed by a
probabilistic projection of the third qudit onto $\ket{0}^\perp$,
followed by the inverse controlled-$X$ gates. If we use two controlled-$X^{-1}$
gates controlled by the first two qudits respectively, the projection
will remove the components with labels $\ket{0}\ket{0}\ket{0}$, $\ket{1}
\ket{0}\ket{1}$ and $\ket{0}\ket{1}\ket{1}$. In addition, using two 
controlled-$X^{(d-1)/2}$ gates, we remove $\ket{1}\ket{1}\ket{1}$ and 
$\ket{0}\ket{0}\ket{0}$ (again). These are the four states 
that need to be removed to produce the ancilla $\ket{\Phi'}$.

The above two procedures allow us to finally produce the desired magic states.
Starting with $\ket{\tilde 0}\otimes\ket{\tilde 0}$, we apply the first
procedure to each ancilla and then apply the second procedure to the appended
qudits. The resulting state is
\beq
\frac{1}{d} \sum_{i=0}^{d-1} \sum_{j=0}^{d-1} \ket{i}\otimes\ket{j}\otimes
\ket{\delta_{i,0}}\otimes\ket{\delta_{j,0}}\otimes
\ket{\delta_{i,0}\delta_{j,0}}.
\eeq

\noindent
If the last three qudits are measured in the $X$ basis, and the results are 
$0$, $0$ and $1$ respectively, then we will have produced
the magic state $\ket{\phi_{M2}}$.

In fact, measuring in the $X$ basis and only accepting if the result is zero
is a convenient way to unentangle the system with temporary qudits.
Therefore, the previously described procedures can be combined into the
probabilistic transformation
\beq
\sum_{i=0}^{d-1}\sum_{j=0}^{d-1} \psi_{i,j} \ket{i}\otimes\ket{j}
\longrightarrow 
\sum_{i=0}^{d-1}\sum_{j=0}^{d-1} \psi_{i,j} \ket{i}\otimes\ket{j}
\otimes \ket{\delta_{i,n}\delta_{j,m}},
\eeq

\noindent
where the first state is either transformed into the second state with some
nonzero probability, or else it is damaged. The above transformation has
only been discussed so far for $n=m=0$, but a trivial use of $X$ gates
before and after the transformation will allow any $n$ and $m$.

Starting with $\ket{\tilde 0}\otimes\ket{\tilde 0}\otimes\ket{0}$,
repeated application of the above procedure can produce
\be
&&\!\!\!\!\!\!\!\!\!\!
\frac{1}{d} \sum_{i=0}^{d-1}\sum_{j=0}^{d-1} \ket{i}\otimes\ket{j}
\otimes\ket{0}
\nonumber\\
&&\!\!\!\!\!\!\!\!\!\!\longrightarrow
\frac{1}{d} \sum_{i=0}^{d-1}\sum_{j=0}^{d-1} \ket{i}\otimes\ket{j}
\otimes\ket{0}  
\bigotimes_{n=0}^{d-1} \bigotimes_{m=0}^{d-1} \ket{\delta_{i,n}\delta_{j,m}}
\nonumber\\
&&\!\!\!\!\!\!\!\!\!\!\longrightarrow
\frac{1}{d} \sum_{i=0}^{d-1}\sum_{j=0}^{d-1} \ket{i}\otimes\ket{j}
\otimes\ket{i j}  
\bigotimes_{n=0}^{d-1} \bigotimes_{m=0}^{d-1} \ket{\delta_{i,n}\delta_{j,m}},
\ee

\noindent
where the second step involves only controlled-$X$ gates from the extra
qudits to the third qudit. Erasing the extra qudits with a measurement in
the $X$ basis, and retaining only when all results are zero, produces
the desired magic state $\ket{\phi_{M1}}$.

The construction of the magic states out of the probabilistic projection onto 
$\ket{0}^\perp$ completes the description of the Toffoli gate. Though the 
procedures of this section are far from optimal in terms of resources, 
they are sufficient to demonstrate universality. In particular, this completes
the proof that universal quantum computation is feasible with anyons from
groups of the form $\Z_p \sdp{\theta} \Z_q$.

\section{\label{sec:gengrp}Computational power of magnetic charges}

In this section, we will be interested in classifying the computational power
that can be achieved by braiding anyonic magnetic charges of a finite
group. The range of operations that can be achieved by braiding is closely 
related to the structure of the group to which the magnetic charges belong.
In particular, the possibility of realizing the operations of controlled-$X$
and Toffoli (equivalently a doubly-controlled-$X$) are respectively related 
to the group properties of nilpotency and solvability. These standard 
properties of group theory will also be defined below.

There are certain important assumptions that go into the discussion in
this section. First, we assume that each qubit is carried by a pair
of anyons. Furthermore, we choose a computational basis corresponding to
the states of definite flux (e.g., $\ket{0}=\ket{g}$ for some $g\in G$).
We remind the reader at this point that the state $\ket{g}$ corresponds
to an anyon of magnetic charge $g$ paired with a compensating anyon of charge
 $g^{-1}$ whose only purpose is to allow the pair to move through the system 
without introducing undesired correlations. Finally, we will restrict the 
discussion to operations that can be achieved by braiding magnetic charges.
The consequences of lifting these restrictions will be discussed near
the end of this section.

Let the fluxes corresponding to the zero and one states be the elements 
$b,\, b'\in G$ respectively. If we desire a coherent 
superposition between the zero and one states, they must be in the same 
conjugacy class, and therefore $b' = a b a^{-1}$ for some non-trivial 
$a\in G$. This is summarized by:
\be
\ket{0} = \ket{b},\ \ \ \   \ket{1} = \ket{b'} = \ket{a b a^{-1}}.
\ee

\noindent
Even if the basis in use is a qudit basis, with additional states,
we will only concern ourselves with states that have support on the above
two basis vectors.

Consider now a pair of these states. We are interested in the operations 
that can be achieved by conjugating the second state by the flux of the first 
state, with the help of ancillas. Let $g\in\{b,b'\}$ be the flux of the 
first state. The most general conjugation possible is by a function of the form
\be
f(g) &=& c_1 g c_2 g c_3 g \cdots g c_n
\\\nonumber
&=&
    \lp( d_1 g' d_1^{-1} \rp)
    \lp( d_2 g' d_2^{-1} \rp)\cdots 
    \lp( d_{n-1} g' d_{n-1}^{-1} \rp) d_n,
\ee

\noindent
for some fixed elements $\{c_i\}\in G$. In the second line, the expression has
been rewritten in terms of $g'=g b^{-1}$, and new elements $\{d_i\}\in G$
which can easily be determined in terms of $\{c_i\}\in G$. For example,
$d_2 = c_1 b c_2$.

The power of the second line is that it expresses the conjugation as a 
composition of two basic operations. The first is a conjugation by an
ancilla with flux $d_n$, and is independent of the state of the first 
qubit. The second is conjugation by a product of conjugates of $g'$, which 
was defined so that if $g=b$ then $g'=1$ and the product of its conjugates is 
trivial.
In the other case, if $g=b'$ then $g'=[a,b]\equiv a b a^{-1} b^{-1}$, and the 
operation is conjugation by a product of conjugates of $[a,b]$.

We define $\mathcal{C}_G(x)$ as the conjugacy class of $x$ in $G$, and
$\mathcal{C}_G^\#(x)$ as the group generated by the elements in 
$\mathcal{C}_G(x)$. The operations discussed so far are 
conjugation by fixed elements in $G$, and controlled conjugation by elements
in $\mathcal{C}_G^\#([a,b])$. 

The most natural controlled operation is the
logical controlled-$X$ gate, which acts as a controlled conjugation by $a$. 
Naturally, if $a^2\neq 1$, then we could arrive at the qudit state
$\ket{2}=\ket{a^2b a^{-2}}$. However, our interest lies 
in proving that certain groups cannot produce a controlled-$X$, in which
case it is sufficient to prove that a controlled conjugation by $a$ is 
unfeasible.

It seems that a requirement for a controlled conjugation by $a$ is the
existence of elements $a,b$ such that $a\in \mathcal{C}_G^\#([a,b])$.
There is a potential loophole in the argument, though, because 
different qubits could use different basis fluxes. The target qubit could
use $b_2$ as the zero state, and $a_2 b_2 a_2^{-1}$ as the one state. If
$a_2 \in \mathcal{C}_G^\#([a_1,b_1])$ then the controlled-$X$ would be 
possible.
Considering many qubits requires a sequence of non-trivial elements $\{a_i\}$
and $\{b_i\}$ which satisfy, at a minimum, the conditions:
\be
a_{i+1} \in \mathcal{C}_G^\#([a_i,b_i]).
\ee

The above equations are related to the series of subgroups of $G$, defined by
\be
G^{((j+1))} = \lp [ G^{((j))}, G \rp ],
\ee

\noindent
with base case $G^{((0))}=G$. By definition, if $a_i\in G^{((j))}$ 
then $[ a_i, b_i ] \in G^{((j+1))}$. Furthermore,
since the group $G^{((j+1))}$ is normal in $G$, the requirement
on $a_{i+1}$ reads
\be
a_{i+1} \in \mathcal{C}_G^\#\lp(\lp[ a_i, b_i \rp]\rp) 
\subset G^{((j+1))}.
\ee

\noindent
Of course, $a_1\in G^{((0))} = G$. Therefore, repeating the above argument 
shows that a controlled-$X$ requires $a_{i} \in G^{((i-1))}$ with 
$a_{i}\neq 1$ for every $i\geq 1$.

Given that $G$ is finite, and $G^{((j+1))} \subset G^{((j))}$, the
series must converge after a finite number of subgroups to some
final subgroup $G^{((\infty))}$. The final subgroup can either be trivial or
non-trivial. The groups with $G^{((\infty))}=\{1\}$ are called nilpotent.
The conclusion thus far is that nilpotent
groups cannot implement a controlled-$X$ by braiding. The inverse of this
statement---i.e., that groups that are not nilpotent can implement
the controlled-$X$---will be shown in Section~\ref{sec:meat}.

\subsection{Conjugations with multiple sources}

A similar analysis can be used to study the relationship between 
group structure and gates produced using multiple qubits as sources of 
conjugation. Clearly any group that is not nilpotent can produce a
series of controlled-$X$ gates with different sources. However, certain
groups are capable of producing much more powerful gates such as the 
Toffoli, which is universal for classical computation.

In the rest of this section we shall prove that groups that are solvable
cannot produce a Toffoli gate, or equivalently universal classical 
computation, by braiding magnetic charges.
This connection between universality for classical computation and 
non-solvability had been previously identified by Barrington 
\cite{Barrington:1989} in 1989. Though we shall mostly be interested in
groups that are solvable, this result will place
limits on the power that we can expect to obtain from braiding magnetic
charges.

Just as above, the most general conjugation with $m$ sources is the 
conjugation by a function of the form
\be
f(g_1,\dots,g_m) &=&
    \lp( d_1 g'_{i_1} d_1^{-1} \rp)
    \lp( d_2 g'_{i_2} d_2^{-1} \rp)\cdots 
\nonumber\\
    && \cdots \lp( d_{n-1} g'_{i_{n-1}} d_{n-1}^{-1} \rp) d_n,
\label{eq:multiprod}
\ee

\noindent
where $g'_{i}=g_{i}b^{-1}$ and the indices $i$ take values from $1$ to $m$.
For brevity, we assume that all qubits are expressed in the same basis, 
though the general case would not be very different.

The Toffoli gate is simply a conjugation by a function $f_T(g'_1,g'_2)$,
such that
\be
f_T(c^k, c^l) = a^{k l},
\ee

\noindent
which has been expressed as a function of $g'_i$ and where we introduced
$c\equiv [a,b]$. In order to produce the Toffoli gate using conjugation 
alone, we must be able to express the above equation in the form of 
Eq.~(\ref{eq:multiprod}) with $m=2$. We shall show that this is not possible 
for a solvable group.

For $m=2$, Eq.~(\ref{eq:multiprod}) is a product of conjugates of $g'_1$
and $g'_2$. We can rewrite it by moving all the conjugates of $g'_1$ to the
left, and all the conjugates of $g'_2$ to the right. In the center we will
pick up factors of the form $[d_i g'_1 d_i^{-1}, d_j g'_2 d_j^{-1}]$
and commutators of commutators, and so on. In the end we will obtain:
\be
f(g'_1, g'_2) = f_1(g'_1) f_C(g'_1, g'_2) f_2(g'_2) d_n,
\ee

\noindent
where $d_n$ is a constant element of $G$, $f_i$ is a product of conjugates 
of $g'_i$, and $f_C(g'_1, g'_2)$ is the factor with all the commutators.
The function $f_C$ has the property that $f_C(g_1',1)=f_C(1,g_2')=1$.

Setting $f=f_T$ implies the conditions
\be
1 &=& f(1,1) = d_n,
\nonumber\\
1 &=& f(c,1) = f_1(c) d_n,
\nonumber\\
1 &=& f(1,c) = f_2(c) d_n,
\nonumber\\
a &=& f(c,c) = f_1(c) f_C(c,c) f_2(c) d_n,
\ee

\noindent
which imply $f_C(c,c)=a$. 

However, $f_C$ has the additional property that, if $N$ is a normal subgroup 
of $G$ containing $c$, then $f_C(c,c)\in [N,N]$. Furthermore, since 
$c=[a,b]$, the requirement on $c$ needed to express the Toffoli function 
in product form is
\be
c\in N \Longrightarrow c\in[N,N],
\label{eq:nosolv}
\ee

\noindent
for any normal subgroup $N$. This condition is related to the 
series of subgroups defined by
\be
G^{(j+1)} = \lp [ G^{(j)}, G^{(j)} \rp ],
\ee

\noindent
again with base case $G^{(0)}=G$. Just as before, this series must
converge to a final subgroup $G^{(\infty)}$. The groups where $G^{(\infty)}=
\{ 1 \}$ are known as solvable. Any group that is nilpotent is also solvable.

Because the subgroups $G^{(j)}$ are all normal in $G$, 
the requirement of Eq.~(\ref{eq:nosolv}) can only be satisfied if
$c$, which by definition cannot be 1, is contained in $G^{(\infty)}$. 
We have therefore shown that if the group is solvable, then the function
$f_T$ cannot be expressed in product form, and therefore we cannot conjugate 
by it. This is true even if the target of conjugation is in a known state, 
which implies that even if we had used the target as a source of conjugations 
as well (i.e., by using it to conjugate ancillas, and then using the ancillas) 
the Toffoli gate would still not be feasible by using only braiding of
anyons from a solvable group.

The fact that the Toffoli gate can be produced for non-solvable groups
is a consequence of the results of Refs.~\cite{me2002,Barrington:1989} 
and will not be discussed here.
In fact, the computational model discussed in this section resembles
the non-uniform deterministic finite automata presented in
Ref.~\cite{Barrington:1990}. For non-solvable groups, the two models
are almost identical. Nonetheless, for solvable groups, the magnetic
charges presented in this section have significantly less computational
power, because the zero and one states have to be represented by 
group elements in the same conjugacy class.

\subsection{Summary of computational power}

The results discussed so far have been summarized by Table~\ref{Tbl:Groups}.
For each type of group, it describes the computational operations that can
be achieved through braiding of magnetic fluxes, as well as an example.
The examples are the smallest group in the class, with the exception of
the abelian case where the trivial group could also be listed. For the
non-abelian, nilpotent case there are two examples with eight elements:
the dihedral group $D_4$, and the quaternionic group $\Q$, which is listed
in the table and has elements $\pm 1, \pm i, \pm j, \pm k$.

\begin{table}
\begin{tabular}{|c|c|c|c|c|}
\hline
Abelian & Nilpotent & Solvable & Example & Computational Power\cr
\hline
  yes   &    yes    &   yes    & $\Z_2$  & $I$ \cr
  no    &    yes    &   yes    & $\Q$     & $X$ \cr
  no    &    no     &   yes    & $S_3$   & Controlled-$X$ \cr
  no    &    no     &   no     & $A_5$   & Toffoli \cr
\hline
\end{tabular}
\caption{Computational power achieved by conjugation for different groups.}
\label{Tbl:Groups}
\end{table}

The most basic case is when $G$ is abelian, in which case it is also
nilpotent and solvable. Clearly conjugation can only produce the identity
transformation. In fact, every superselection sector consists of a one
dimensional Hilbert space, and therefore quantum information cannot
even be stored in abelian anyons in a topologically protected manner.

At the other extreme are anyons from non-solvable groups. Universal classical
computation can be accomplished through braiding, and universal quantum
computation can be obtained by completing the gate-set with measurements
in the $X$ and $Z$ bases. The complete construction for this case is
described in Ref.~\cite{me2002}.

Anyons from groups that are solvable but not nilpotent, can also be used
for universal quantum computation, but the construction is more complicated.
A controlled-$X$ can be constructed from flux braiding, and measurements in 
the $X$ and $Z$ bases can be constructed in a manner similar to the 
non-solvable groups. However, to complete a universal gate-set, fusions 
of electric charges must be employed. The proof of universality, 
along with the details of the gates, will be the subject of the rest of 
this paper.

Finally, anyons from groups that are nilpotent seem insufficient for 
universal computation. In the constructions for the non-nilpotent groups,
the only operation that can produce entanglement between multiple qudits
is the controlled-$X$ or Toffoli gates obtained by braiding fluxes.
However, for nilpotent groups, braiding fluxes does not seem to yield
an operation capable of producing entanglement. Either a new type
of operation, or a different basis must be used. Simple modifications to
the basis, such as encoding a qudit on multiple anyons, are of no
help. However, there are countless strange bases that are hard
to discredit. For example, a lattice of electric charges could serve
as a Hilbert space, with magnetic charges used to create or measure 
entanglement among the charges. Therefore, while the prospects of universal
computation with nilpotent anyons seem bleak, the question remains open.

\section{\label{sec:meat}Solvable non-nilpotent groups}

In this section, we will prove that anyons based on a finite group that is 
solvable but not nilpotent are sufficient for universal quantum computation. 
The first step will be to decompose an arbitrary group $G$ that is solvable
but non-nilpotent, into a form similar to the previously studied 
$\Z_p \sdp{\theta} \Z_q$ groups. The proof of universality will then be
a small generalization of the ideas presented in Section~\ref{sec:sdp}.

\subsection{Group decomposition}

Let $G$ be a group as above and define $H\equiv G^{((\infty))}$ in terms
of the series discussed in Section~\ref{sec:gengrp}. Because $G$ is 
non-nilpotent $H$ is non-trivial, and because $G$ is solvable $H\neq G$. 
Furthermore, $H$ is normal in $G$ and $G/H$ is nilpotent. The second fact
is due to
\be
\lp( G/H \rp)^{((i+1))} &=& \lp[ \lp( G/H \rp)^{((i))}, \lp( G/H \rp) \rp]
\nonumber\\
&=& \lp[ G^{((i))}, G \rp ]/H = G^{((i+1))}/H,
\ee

\noindent
and therefore $( G/H )^{((\infty))}=H/H=\{1\}$. 

Any nilpotent group
can be written as the direct product of its Sylow p-groups, which are
groups whose order is a prime power. Therefore
\be
G/H = K_{q_1} \times K_{q_2} \times \dots \times K_{q_l},
\ee

\noindent
where $K_{q}$ denotes a group of order $q^m$ for some prime $q$ and integer 
$m$. We further define $N K_{q_i}$ to be the lifting of $K_{q_i}$ to
the full group $G$ that is $N K_{q_i}/N=K_{q_i}$.
Note that to maintain consistency with the notation in Section~\ref{sec:sdp},
the primes involved in these p-groups are labeled by the letter $q$.

Having fully characterized $G/H$, we turn to the study of $H$ itself.
Let $N$ be the largest normal subgroup of $G$ that also satisfies
$N\subset H$ and $N\neq H$.
If more than one subgroup satisfies the above requirements, then
let $N$ be any such subgroup. Because $H$ is finite, there must be at least
one maximal subgroup.

We shall prove that $H/N= \Z_p^n$ for some prime $p$ and integer $n$.
The basic idea is that working modulo $N$, $H/N$ is a normal subgroup
of $G/N$. Furthermore $H/N$ has no proper subgroups that are normal
in $G/N$. In particular, this implies that $H/N$ is abelian, because
its commutator subgroup is a normal subgroup of $G/N$. Note that
the possibility that the commutator subgroup of $H/N$ is equal to
$H/N$ is excluded because $H/N$ is solvable.

For any $x\in H/N$ consider $\mathcal{C}_{G/N}^\#(x)$, the group generated
by the conjugates of $x$ in $G/N$. This is a subgroup of $H/N$ and 
is normal in $G/N$. Therefore
\be
\mathcal{C}_{G/N}^\#(x) = H/N,\ \ \ \ \forall x\in H/N,
\ee

\noindent
which implies that all elements in $H/N$, with the exception of the identity,
have the same order. That is because conjugates of $x$ have the same order
as $x$, and a product of elements of order $k$, in an abelian group,
must have order less than or equal to $k$. This concludes the proof
that $H/N= \Z_p^n$.

Thus far, we have the following tower of groups
\be
N \subset H \subset HK_{q_i} \subset G,
\ee

\noindent
where $N$, $H$ and $HK_{q_i}$ are all normal in $G$, and the group 
$HK_{q_i}$ can be any of the groups found above. 

Because $(G/N)^{((\infty))}=H/N=\Z_p^n$, the group $G/N$ is also solvable
and non-nilpotent. However, its structure is simpler than that of the full 
group $G$. We shall therefore be interested in working modulo $N$, and shall
denote groups modulo $N$ by a tilde. That is
\beq
\tilde G = G/N, \ \ \ \tilde H K_{q_i} = H K_{q_i}/N, \ \ \ 
\tilde H = H/N = \Z_p^n.
\eeq

The final step is to study the relationship between $\tilde H$ and the groups
$\tilde H K_{q_i}$. By construction, we know 
$[\tilde G, \tilde H] = \tilde H = Z_p^n$, but what about 
$[\tilde H K_{q_i}, \tilde H]$?  
Because both $\tilde H K_{q_i}$ (for any $i$) and $\tilde H$ are normal 
in $\tilde G$,  $[\tilde H K_{q_i}, \tilde H]$ is normal in $\tilde G$ and,
furthermore, is contained in $\tilde H$. But $N$ was defined to be the largest
proper subgroup of $H$ that was normal in $\tilde G$. Therefore $\tilde H$
has no proper subgroups that are normal in $\tilde G$, and 
$[\tilde H K_{q_i}, \tilde H]$ must be either the trivial group or 
all of $\tilde H$.

If $q_i=p$ then $\tilde H K_{q_i}$ is a p-group, and therefore nilpotent.
This means that $[\tilde H K_{p}, \tilde H]\neq \tilde H$ and by the
previous paragraph $[\tilde H K_{p}, \tilde H] = \{1\}$. The rest of the
groups $\tilde H K_{q_i}$ can either commute or not with $\tilde H$. However,
because $[\tilde G,\tilde H]=\tilde H$, at least one of them must not commute.
Fix an $i$ such that $[\tilde H K_{q_i}, \tilde H]= \tilde H$, and define
$K = K_{q_i}$, $\tilde HK = \tilde HK_{q_i}$ and $q=q_i$. 
This will be the group to take the place of $\Z_q$.

We would like to show that there exists an element $b\in \tilde H K$, such that
$[b,\tilde H]=\tilde H$. Let $X$ be the stabilizer of $\tilde H$ in 
$\tilde H K$, that is, the largest subgroup of $\tilde H K$ such
that $[X,\tilde H]=1$. Clearly $H \subset X$ and $X\neq \tilde H K$.
Because $\tilde H K/ X$ is nilpotent, it has a non-trivial center. Let
$b\in \tilde H K$ be any element that projects, modulo $X$ to one of the
non-trivial elements in the center. We will show that $[b, \tilde H]$ is
normal in $\tilde G$, which implies $[b, \tilde H]=\tilde H$. The proof is that
modulo $X$ (which is normal in $\tilde G$), every element $g\in\tilde G$
commutes with $b$. Therefore $g b g^{-1}= b x$ for some $x\in X$ and
\be
g\lp[b,h \rp] g^{-1} 
&=& b x h' x^{-1} b^{-1} h'^{-1} 
\nonumber\\
&=& b h' b^{-1} h'^{-1} \in \lp[b, \tilde H \rp],
\ee

\noindent
for any $h\in \tilde H$, where $h'=g h g^{-1}\in \tilde H$.

To summarize, working modulo $N$, we have the following tower of subgroups:
\be
\tilde H \subset \tilde H K \subset \tilde G,
\ee

\noindent
with $\tilde H = Z_p^n$ for some prime $p$. Furthermore, 
$\tilde H K / \tilde H = K$ 
is a subgroup of order a power of $q$, for some prime 
$q$ not equal to $p$. Finally, $\exists b \in \tilde H K$ such that
$[b, \tilde H] = \tilde H$.

Note that this notation is consistent with the one used in 
Section~\ref{sec:sdp}. That is, if $G = \Z_p \sdp{\theta} \Z_q$, then
$N=\{1\}$, $H=\Z_p$, $K=\Z_q$, and the definitions of $p$, $q$, and $b$
would be consistent.

\subsection{Examples}

There are a few good examples to keep in mind that illustrate the potential
new complications arising from groups with more structure than 
$\Z_p \sdp{\theta} \Z_q$.

The first example is $A_4 = \Z_2^2 \sdp{\theta} \Z_3$. The group can be 
described as $a_1 = (12)(34)$, $a_2 = (23)(41)$, $b=(123)$ with 
\be
a_1^2 = a_2^2 = 1, & \ \ \ & a_1 a_2 = a_2 a_1, \nonumber \\
b^3 = 1, & \ \ \  & b a_1^i a_2^j b^{-1} = a_1^j a_2^{i+j}.
\ee

\noindent
For this group $N=\{1\}$, $H=\Z_2^2$ and $K=\Z_3$. Its most important feature 
is that $p=2$, which was not previously possible. Because $p=2$ implies working
with qubits, these groups will be have to be handled specially.

The next example is $G=(\Z_3^2) \sdp{\theta} (\Z_3 \times \Z_2)$.
Let $a_1$, $a_2$ be the generators of $\Z_3^2$, let $b$ be the generator of 
$\Z_2$ and $x$ be the generator of the remaining $\Z_3$. The semidirect
product is defined by the conjugations
\be
b a_1^i a_2^j b^{-1} = a_1^{-i} a_2^{-j}, \ \ \ 
x a_1^i a_2^j x^{-1} = a_1^{-j} a_2^{i-j}.
\ee

\noindent
For this group $H=\Z_3^2$ because $[G,G]=[G,H]=H$. The subgroup
generated by $a_1 a_2^{-1}$ is normal in $G$ and therefore $N=\Z_3$.
Finally $H/N=\Z_3$ and $K = \Z_2$. Note that $x$ commutes with 
$H$ modulo $N$, as discussed in the last section.

The final pair of examples illustrate the case where $K$ is non-abelian. 
The examples are $\Z_3^2 \sdp{\theta} \Q$ and $\Z_3^2 \sdp{\theta} D_4$.
Labeling the generators of $\Z_3^2$ by $a_1$ and $a_2$, the semidirect
product for $\Z_3^2 \sdp{\theta} \Q$ is defined by
\be
i a_1^x a_2^y i^{-1} = a_1^{y} a_2^{-x},
\ \ \ \ \ 
j a_1^x a_2^y j^{-1} = a_1^{x+y} a_2^{x-y},
\ee

\noindent
where $\pm 1$, $\pm i$, $\pm j$, $\pm k$
are the standard quaternionic elements. For $\Z_3^2 \sdp{\theta} D_4$ the
semidirect product is defined by
\be
\beta a_1^x a_2^y \beta^{-1} = a_1^{y} a_2^{-x},
\ \ \ \ \ 
\gamma a_1^x a_2^y \gamma^{-1} = a_1^{y} a_2^{x},
\ee

\noindent
where the relations $\beta^4=\gamma^2=1$ and $\gamma\beta\gamma=\beta^{-1}$
define $D_4$.

In both of the above cases $p=3$, $q=2$, $N=\{1\}$ and $H=\Z_3^2$. 
However, for $\Z_3^2 \sdp{\theta} \Q$ the non-trivial elements of $H$ are 
conjugate to one another, and none of the non-trivial elements of $Q$ commute 
with any of the non-trivial elements of $H$. 
The $\Z_3^2 \sdp{\theta} D_4$ case divides $H$ into three conjugacy 
classes (including the identity). Furthermore, each of the elements of the 
form $\beta^i \gamma$ commute with two non-trivial elements of $H$. 
These differences will become important
when discussing the operations involving electric charges.

\subsection{\label{sec:newreq}$N$-invariant ancillas}

The first lesson from the above analysis is that we should work modulo $N$.
That is, we want flux states labeled by elements of $\tilde G=G/N$,
that are invariant under $N$. The idea of $N$-invariant states was already
discussed in Ref.~\cite{me2002} when generalizing simple non-abelian anyons
to non-solvable ones, and therefore the discussion below will be brief.

A basis for the $N$-invariant magnetic fluxes is just $\ket{g}$ for 
$g\in\tilde G$. The braiding and fusion properties of these states behave
almost exactly as if the full group were $\tilde G$ and these states were flux 
eigenstates. The only difference is that when fusing two anyons from pairs
with opposite fluxes, the probability of disappearing into the vacuum is lower.

Even producing anyons from the vacuum behaves correctly with respect to 
$N$-invariance. Pairs produced from the vacuum are naturally invariant
under the full group $G$. Normally, when braiding with other states, this 
invariance will be broken. However, if the vacuum pair only interacts with
$N$-invariant states, then the invariance under the group $N$ will remain.

At this point we will change our requirements for the physical system. 
Instead of requiring a reservoir of flux ancillas for every element of $G$,
we will require a reservoir of $N$-invariant flux ancillas for every 
element of $\tilde G$. This is likely a reasonable modification, as it appears
that the latter ancillas are no harder to produce than the original ones.

It should be noted that, when working modulo $N$, the electric charges need no 
modification. That is, because $N$ is normal in $G$, any representation of
$\tilde G$ extends to a representation of $G$ that is invariant under $N$.
Furthermore, fusing two $N$-invariant electric charges must produce a new
$N$-invariant electric charge. Therefore, working with $N$-invariant electric
charges simply involves working with a subset of the charges of the group $G$.

Given the above caveats, we can effectively replace the group $G$ with the
group $\tilde G=G/N$, which will be done without further comment for the rest 
of this section.

\subsection{Computational basis}

We will begin by defining an extended computational basis, and discuss the
operations that can be performed on this extended subspace. Toward the end
of this section, a subset of these states will be singled out as the
true computational basis.

Let $a_1,\dots,a_n$ be a set of generators for $\tilde H=\Z_p^n$, and recall
the definition of the element $b\in \tilde G$. The extended computational
basis consists of the states
\be
\ket{i_1,\dots,i_n} \equiv \ket{a_1^{i_1}\cdots a_n^{i_n} b 
a_n^{-i_n}\cdots a_1^{-i_1}},
\ee

\noindent
where each of the $i$'s takes values from $0$ to $p-1$. 

To prove that the states are all distinct consider the 
map from $H\rightarrow H$ defined by
\be
\lp[ g, \cdot \rp ]: h \rightarrow \lp[ g, h \rp ].
\ee

\noindent
Because $\tilde H$ is abelian, this map
is an homomorphism for any $g\in\tilde G$. In particular, since 
$[b,\tilde H] = \tilde H$, the homomorphism defined by $[b,\cdot]$ is
surjective and has trivial kernel. That is, no element of $\tilde H$ commutes
with $b$. But
\beq
h b h^{-1} = h' b h'^{-1} \Rightarrow (h'^{-1} h)b (h'^{-1} h)^{-1} = b,
\eeq

\noindent
for any elements $h,h'\in \tilde H$, which can only be true if $h=h'$.

\subsection{Basic operations}

The generalized controlled-$X$ is the transformation
\be
&&\ket{i_1,\dots,i_n}\otimes\ket{j_1,\dots,j_n}
\nonumber\\
&&\longrightarrow
\ket{i_1,\dots,i_n}\otimes\ket{i_1+j_1,\dots,i_n+j_n}.
\ee

\noindent
It can be implemented as a conjugation of the second anyon by a function
of the flux of the first anyon such that
\be
f(h b h^{-1}) = h,
\ee

\noindent
for any $h\in\tilde H$. 
Because the map $[b,\cdot]$ defined above is just a permutation of the 
elements of $\tilde H$, it has a finite period (say $l$). The desired function
is
\be
f(g) = \bigg [ \Big [ \big [ g b^{-1}, b \big ],b \Big]\cdots,b
\bigg ],
\label{eq:commuts}
\ee

\noindent
which consists of $l-1$ nested commutators. The final commutator needed
to complete the period is the one formed in the expression $g b^{-1}$ when
$g$ has the form $h b h^{-1}$.

At this point, one may wonder how does working modulo a normal subgroup $N$ 
affect the discussion regarding the computability of the controlled-$X$ gate. 
The controlled-$X$ can only be implemented because $\tilde G$ is non-nilpotent.
In a sense, $\tilde G$ was constructed to be as small as possible, but still
maintain the property of being non-nilpotent. On the other hand, if a group
$G$ is nilpotent to begin with, then any subgroup or quotient group
will also be nilpotent, and no controlled-$X$ can be constructed 
using braiding.

Using the same techniques as in Section~\ref{sec:fuse}, anyon fusions can
be used to perform measurements. Fusion with $\ket{b^{-1}}$ ancillas
produces a probabilistic projection onto $\ket{0,\dots,0}$. Fusing the two
anyons that form a qudit is a probabilistic projection onto
\be
\ket{\tilde 0,\dots,\tilde 0} \equiv \frac{1}{\sqrt{p^n}}
\sum_{i_1=0}^{p-1} \cdots \sum_{i_n=0}^{p-1} \ket{i_1,\dots,i_n}.
\ee

As usual, to complete the probabilistic projection, these fusions must be 
supplemented by a reservoir of $\ket{0,\dots,0}$ and 
$\ket{\tilde 0,\dots,\tilde 0}$ ancillas. The first case is trivial, because
the existence of these ancillas has been assumed as one of the physical 
requirements of the system. The production of $\ket{\tilde 0,\dots,\tilde 0}$ 
ancillas is more complicated and will occupy the rest of the subsection.

The procedure to distill $\ket{\tilde 0,\dots,\tilde 0}$ states begins
with a pair created from the vacuum and a $\ket{0,\dots,0}$ ancilla:
\be
\ket{\mbox{Vac}}\otimes\ket{0,\dots,0}.
\ee

\noindent
Using only braiding, an incomplete swap is applied to the state:

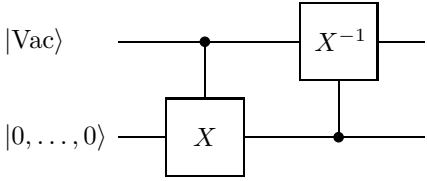
\begin{center}
\setlength{\unitlength}{1in}
\begin{picture}(2.8,1.1)(-.2,0)
{\put(.7,.8){\line( 1, 0){.95}}}
{\put(2.3,.8){\line( -1, 0){.25}}}
{\put(.7,.3){\line( 1, 0){.25}}}
{\put(2.3,.3){\line( -1, 0){.95}}}
{\put(1.15,.8){\line( 0,-1){.3}}}
{\thinlines \put(1.15,.8){\circle*{.05}}}
{\put(0.95,.1){\framebox(.4,.4){}}}
{\put(1.85,.3){\line( 0,1){.3}}}
{\thinlines \put(1.85,.3){\circle*{.05}}}
{\put(1.65,.6){\framebox(.4,.4){}}}
\put(1.15,.3){\makebox(0,0){$X$}}
\put(1.86,.81){\makebox(0,0){$X^{-1}$}}
\put(.1,.8){\makebox(0,0)[l]{$\ket{\mbox{Vac}}$}}
\put(.1,.3){\makebox(0,0)[l]{$\ket{0,\dots,0}$}}
\end{picture}
\end{center}

\noindent
Once again, the circuit denotes the action of the conjugations on the 
computational basis, but their extension to the full Hilbert space needs to
be discussed. After applying the necessary braidings to perform the circuit, 
the top state is fused with a $\ket{b^{-1}}$ ancilla. If the fusion does not
produce the vacuum state, the final product is discarded and the procedure
restarted from the beginning. Since conjugations
cannot change the superselection sector, the only case that needs to be 
considered is when the vacuum state is created in the superselection 
sector that contains the computational subspace (i.e., the conjugacy
class of $b$). In this superselection sector the vacuum state has the form
\be
\ket{\mbox{Vac}} \propto \ket{\tilde 0,\dots,\tilde 0} + \ket{\Psi^\perp},
\ee

\noindent
where $\ket{\Psi^\perp}$ is a state in the space spanned by vectors of the form
$\ket{g b g^{-1}}$ that are not contained in the computational basis.

Because we want to guarantee that after the controlled-$X$ the state
$\ket{0,\dots,0}$ remains in the computational subspace, we need the
conjugation function to satisfy
\be
f(\tilde G) &\in& \tilde H, \ \ \ \mbox{and}\ \nonumber\\
f(h b h^{-1})&=&h \  \ \ \ \forall h\in \tilde H.
\ee

\noindent
The second requirement can be satisfied by choosing $f$ as a sequence of
commutators as in Eq.~(\ref{eq:commuts}), as long as the number of commutators 
is one minus a multiple of $l$ (the period of $[b,\cdot]$). Furthermore, 
the result after $i$ commutators must be contained in $\tilde G^{((i))}$.
Because the series is finite, 
$\tilde G^{((j))}=\tilde G^{((\infty))}=\tilde H$ for some finite $j$,
and the first requirement can also be satisfied by defining $f$ to be
a long enough sequence of commutators. 
Both requirements can be satisfied simultaneously
by correctly choosing the number of commutators in the expression, and this
completes the definition of the first controlled-$X$.

The second controlled-$X$ can be a regular controlled-$X$ because in this
case the control is known to be in the computational subspace. In the end,
the vacuum state will be conjugated by an element of $\tilde H$, and therefore
can only have flux $b$ if it was originally in the computational subspace.

Having completed the construction of the  $\ket{\tilde 0,\dots,\tilde 0}$
ancillas, all that is required to complete a universal
set of gates is an analog of the probabilistic projection onto $\ket{0}^\perp$
constructed out of fusions of electric charges.

\subsection{Using electric charges}

The ideal goal for this section would be the construction of 
the probabilistic projection onto $\ket{0,\dots,0}^\perp$ gate. Unfortunately,
this is not possible for most groups. However, we will produce a pair
of gates that have an equivalent computational power.

The first gate involves a non-trivial subgroup 
$\tilde\Lambda \subset \tilde H$,
to be defined later, which could equal all of $\tilde H$. 
Note that this subgroup defines a subspace of the computational space 
spanned by 
\be
\ket{\lambda b \lambda^{-1}},
\ee

\noindent
for all elements $\lambda\in\tilde\Lambda$, which will also be denoted by 
$\tilde\Lambda$. 
The probabilistic projection onto $\tilde\Lambda$ will be the first gate.

The second gate is the probabilistic projection onto 
$\ket{0,\dots,0}^\perp\cap\tilde\Lambda$. This second gate can be though of
as an application of the first gate, followed by a 
probabilistic projection onto $\ket{0}^\perp$ that only works on states
contained in $\tilde\Lambda$.
For the moment, we will assume that the first gate can be implemented, and 
work on the construction of the second gate.

The basic building block for this section involves working with the state
to be measured $\ket{\Psi}$ and an electric charge pair in the vacuum 
state $\ket{R(I)}_R$ of some non-abelian representation $R$. The state
to be measured is contained in the computational basis and can therefore
be expanded as
\be
\ket{\Psi}=\sum_{h\in \tilde H} \psi_h \ket{h b h^{-1}},
\ee

\noindent
where, as in Section~\ref{sec:Zelec}, the coefficients $\{\psi_h\}$ 
could be numbers or could denote the state of the rest of the system.

Using braiding, the state $\ket{\Psi}$ can be entangled with the electric
charges. In particular, if $\phi(g)$ is a function constructed as a product 
of $g$ and fixed elements of $\tilde G$, then the following transformation 
can be realized:
\be
\ket{\Psi}\otimes\ket{R(I)}_R \longrightarrow
\sum_{h\in \tilde H} \psi_h \ket{h b h^{-1}} \otimes \ket{R(\phi(h))}_R.
\ee

\noindent
Note that the state of the electric charge can depend on $\phi(h)$
rather than $\phi(h b h^{-1})$ by composing with the function defined
in Eq.~(\ref{eq:commuts}). That is $\phi(f(h b h^{-1}))=\phi(h)$.

Now the electric charge pair is fused together, and the resulting particle
is measured. More specifically, in accordance with the discussion in 
Section~\ref{sec:1de}, we just check whether the resulting particle
belongs to some one-dimensional representation labeled $\gamma$.
If the charge $\gamma$ is detected, then the electric charge will
have disentangled with the state being measured, because its internal Hilbert
space is one dimensional. Furthermore, because each one-dimensional 
representation occurs only once in the decomposition of $R\otimes R^*$, 
the state will be unentangled with the environment as well. The proof
of the latter property uses Schur's lemma and the fact that if
$\ket{M_1}_R$ and $\ket{M_2}_R$ always fuse into representation $\gamma$
then $\ket{M_1 M_2^\dagger}_R$ will always fuse into the vacuum.

The result of the complete operation, when the outcome $\gamma$ is obtained,
is the transformation
\be
\ket{\Psi} \longrightarrow
\sum_{h\in \tilde H} F_{\phi(h)\rightarrow\gamma} \psi_h \ket{h b h^{-1}},
\ee

\noindent
where the state after the measurement has been left unnormalized.
The coefficients $F_{h\rightarrow\gamma}$ depend implicitly on the 
original representation $R$, and will be defined carefully below.

The above procedure can be repeated many times for different functions 
$\phi(g)$. If on each occurrence the outcome $\gamma$ is obtained,
the resulting (unnormalized) state will be
\be
\sum_{h\in \tilde H} \lp( \prod_{\phi\in\Phi} F_{\phi(h)\rightarrow\gamma}\rp)
\psi_h \ket{h b h^{-1}},
\ee

\noindent
where $\Phi$ is the set of functions used. As usual, if the outcome $\gamma$
is not obtained on each instance, the state is discarded, and the probabilistic
projection reports a projection onto the complement.

We assume that all functions in the set $\Phi$ are products of conjugates 
of the input, and therefore $\phi(I) = I$ for any $\phi\in\Phi$.
Because $\ket{R(I)}_R$ is the vacuum state, it will always fuse
back into the vacuum. Therefore, if $\gamma$ is a non-trivial representation
then $F_{I\rightarrow\gamma} = 0$ and the above operation projects out the 
$\ket{0,\dots,0}$ state.  

At this point we have almost constructed a probabilistic projection onto
$\ket{0,\dots,0}^\perp \cap\tilde\Lambda$. 
The states outside of $\tilde\Lambda$
can be removed using the probabilistic projection onto $\tilde\Lambda$, which 
for the moment we assume can be implemented. Therefore, the desired 
gate will be complete if the coefficients
\be
\prod_{\phi\in\Phi} F_{\phi(\lambda)\rightarrow\gamma}
\label{eq:coeff}
\ee

\noindent
are non-zero and equal for every non-trivial $\lambda\in\tilde\Lambda$. 
The requirement of equality is accomplished if the orbits
under the functions in $\Phi$, of all non-trivial $\lambda\in\tilde\Lambda$,
are equal.

More specifically, let $\Phi$ be a set of maps from $\tilde\Lambda$ to 
$\tilde\Lambda$ 
that fix the identity. We say that $\Phi$ is balanced on $\tilde\Lambda$ if it
satisfies the relation
\beq
\#(\lambda_1 \rightarrow \lambda') = \#(\lambda_2 \rightarrow \lambda') 
\ \ \ \ 
\forall\  \lambda_1, \lambda_2, \lambda' \in \tilde\Lambda -\{I\},
\eeq

\noindent
where $\#(\lambda\rightarrow\lambda')$ denotes the number of elements 
$\phi\in\Phi$ such that $\phi(\lambda)=\lambda'$. 
The requirement that $\Phi$ be balanced guarantees
that the expressions in Eq.~(\ref{eq:coeff}) are equal for every $\lambda$. 
Of course, for the coefficients to be non-zero, we must prove
separately that the value of $F_{\lambda\rightarrow\gamma}$ is non-zero
for every non-trivial $\lambda\in\tilde\Lambda$.

The goal for the rest of this section is, therefore, to find: a subgroup
$\tilde\Lambda$ of $\tilde H$, an irreducible representation 
$R$ of $\tilde G$, and a one-dimensional representation $\gamma$ of 
$\tilde G$ such that:

\begin{enumerate}
\item The probabilistic projection onto $\tilde\Lambda$ can be implemented.
\item $F_{\lambda\rightarrow\gamma} \neq 0$ for every non-trivial 
$\lambda\in\tilde\Lambda$.
\item There exists a set of maps from $\tilde\Lambda$ to $\tilde\Lambda$,
that is balanced on $\tilde\Lambda$, and can be expressed as
\be
\phi(g) = \prod_i g_i g g_i^{-1},
\ee
\noindent
for some elements $\{g_i\}\in \tilde G$.
\end{enumerate}

\subsubsection{Choosing $\tilde\Lambda$}

There are groups, such as $\Z_3^2\sdp{\theta} D_4$, for which there is no 
choice of $R$ and non-trivial $\gamma$ such that 
$F_{h\rightarrow\gamma}\neq 0$ for all non-trivial $h\in \tilde H$. 
It is therefore advantageous to choose $\tilde\Lambda$ as small as possible.
Furthermore, a small $\tilde\Lambda$ will also help when proving the 
existence of a set of functions balanced on $\tilde\Lambda$.
 
Let $a$ be a non-trivial element of $\tilde H$, and consider the set of
functions of the form 
\be
\phi(g) = \prod_i g_i g g_i^{-1},
\label{eq:conjform}
\ee
 
\noindent
such that $\phi(a)=I$. The kernel of each of these functions is a subgroup of
$\tilde H$ that contains the element $a$. We define $\tilde\Lambda$ as the 
intersection of all these kernels.
 
Because there are a finite set of maps from $\tilde H$ to $\tilde H$, we
can find a finite set of functions $\{\phi_i\}$, in the form of
Eq.~(\ref{eq:conjform}), satisfying
\be
\lambda \in \tilde\Lambda &\Longrightarrow& \forall i\ \ \phi_i(\lambda)=I,
\nonumber\\
h \notin \tilde\Lambda &\Longrightarrow&    \exists i\ \ \phi_i(h)\neq I.
\ee 

\noindent
A probabilistic projection onto $\tilde\Lambda$ can be constructed using 
controlled conjugations on an ancilla $\ket{b}$:
\be
&&\sum_{h\in\tilde H} \alpha_h \ket{h b h^{-1}}\otimes\ket{b}
\nonumber\\
&&\longrightarrow
\sum_{h\in\tilde H} \alpha_h \ket{h b h^{-1}}\otimes
\ket{\phi_i(h) b \phi_i(h)^{-1}},
\ee

\noindent
and then using fusion to make sure that the ancilla remains in the $\ket{b}$
state. Repeating the procedure for each $\phi_i$ produces the desired
projection.

To build the set of functions that are balanced on $\tilde\Lambda$, let 
$\Phi$ be the set of functions in the form of Eq.~(\ref{eq:conjform})
such that $\phi(a)\in\tilde\Lambda -\{I\}$. We shall prove that this is the 
desired set of functions.

Let $\lambda\in\tilde\Lambda$ be non-trivial and let $\phi$ be any map in 
$\Phi$. The value of $\phi(\lambda)$ must be non-trivial and contained in 
$\tilde\Lambda$. Otherwise, it would be possible to construct a map in
product form such that $a$ is in its kernel but $\lambda$ is not, 
contrary to the definition of $\tilde\Lambda$.
In fact, the functions in $\Phi$ are just automorphisms of $\tilde\Lambda$ and
form a group with multiplication given by function composition. Furthermore,
because $\mathcal{C}_{\tilde G}^{\#}(a)=\tilde H$, for any non-trivial
$\lambda\in\tilde\Lambda$ there exists a function 
$\phi_{a\rightarrow\lambda}\in\Phi$ such that 
$\phi_{a\rightarrow\lambda}(a)=\lambda$. 
If $\lambda'\in\tilde\Lambda$ is a third 
non-trivial element, then for every function $\phi\in\Phi$ such 
that $\phi(\lambda)=\lambda'$ there is a function $\phi'(a)=\lambda'$ given by 
$\phi'=\phi\circ\phi_{a\rightarrow\lambda}$. Therefore, $\Phi$ is balanced 
on $\tilde\Lambda$.

\subsubsection{The amplitudes $F_{h\rightarrow\gamma}$}

To choose $R$ and $\gamma$ we first need to examine and define
$F_{h\rightarrow\gamma}$ more carefully. Since we are mostly interested in 
whether $F_{h\rightarrow\gamma}$ is zero or non-zero, we will generally 
work with its magnitude squared, which has the simple expression
\be
\lp|F_{h\rightarrow\gamma}\rp|^2 = 
\Big| P_\gamma \ket{R\lp(h\rp)}_R \Big|^2,
\ee

\noindent
where $P_\gamma$ is the projector onto the space that will turn into
the representation $\gamma$ after fusion. This subspace 
is just the subspace that transforms as $\gamma$ under conjugation.
It can be projected out using the orthogonality of characters 
(and matrix entries for non-abelian representations):
\be
P_\gamma \ket{\Psi} = \frac{1}{|\tilde G|}
\sum_{g\in\tilde G} \bar\gamma^g \ U(g)\otimes U(g) \ket{\Psi},
\ee

\noindent
where $\bar\gamma$ is the conjugate representation. Note that the
values of the representation $\gamma$ on $g\in\tilde G$ will be denoted by 
$\gamma^g$, as a reminder that it is always a power of some root of unity
which we shall also denote by $\gamma$. 

Combining the expressions for the projector and the electric charge state we
obtain
\be
\lp|F_{h\rightarrow\gamma}\rp|^2 &=& 
\bigg| \frac{1}{|\tilde G|}
\sum_{g\in\tilde G} \bar\gamma^g \ket{R\lp(g h g^{-1}\rp)}_R
\bigg|^2
\nonumber\\
&=& \frac{1}{d_R |\tilde G|^2} \bigg|
\sum_{g\in\tilde G} \bar\gamma^g R\lp(g h g^{-1}\rp)
\bigg|^2,
\ee

\noindent
where $d_R$ is the dimension of representation $R$. In the second line,
the magnitude squared of the matrix is given by $|M|^2=\Tr{MM^{\dagger}}$,
which is equivalent to the sum of the magnitude squared of the entries of 
the matrix. 

Because $\tilde H$ is abelian, the
representation $R$ can be diagonalized on $\tilde H$ so that the
diagonal entries are one-dimensional representations of $\tilde H$.
These representations can be labeled by an index $i$ running along the 
diagonal of the matrices $R$, and described by functions 
$\omega_i^h:\tilde H \rightarrow \C$. With the new notation:
\be
\lp|F_{h\rightarrow\gamma}\rp|^2 =\frac{1}{d_R |\tilde G|^2} 
\sum_{i=1}^{d_R} \bigg| 
\sum_{g\in\tilde G} \bar\gamma^g\, \omega_i^{g h g^{-1}}
\bigg|^2,
\label{eq:noS}
\ee

\noindent
where the representation $R$ is now implicit in the definition of the
representations $\{\omega_i\}$.

Finally, let $\tilde S$ be the stabilizer of $\tilde H$ in $\tilde G$,
that is, the subgroup of $G$ that commutes with every element of $\tilde H$.
Clearly, it is a normal subgroup of $\tilde G$, and $\tilde H\subset \tilde S$.
Furthermore, we had argued that if $q_i=p$ then $K_{q_i} \in \tilde S$. 
Therefore $|\tilde G/\tilde S|$ is not divisible by $p$.

Since the function $F_{h\rightarrow\gamma}$ will be zero unless we choose a 
representation such that $\gamma^{\tilde S}=1$, we shall assume this from now
on, and write
\be
\lp|F_{h\rightarrow\gamma}\rp|^2 =\frac{|\tilde S|^2}{d_R |\tilde G|^2} 
\sum_{i=1}^{d_R} \bigg| 
\sum_{g\in\tilde G/\tilde S} \bar\gamma^g\, \omega_i^{g h g^{-1}}
\bigg|^2.
\ee

We are now guaranteed that $\gamma$ corresponds to powers of an $n^{th}$ root
of unity such that $p$ does not divide $n$. The terms in the above expression
have the form
\be
\sum_{i=0}^{p-1} c_i \omega^i,
\label{eq:omegacoeff}
\ee

\noindent
where the coefficients $c_i$ are sums of $n^{th}$ roots of unity. By 
Ref.~\cite{Shoenberg}, the expression will be zero if and only if
the $n$ coefficients $c_i$ are all equal.

Using the above notation it is easy to show two properties of the
amplitudes $F_{h\rightarrow\gamma}$. If $|F_{h\rightarrow\gamma}|\neq 0$ then
\be
\lp|F_{h^j\rightarrow\gamma}\rp|^2 &=&\frac{|\tilde S|^2}{d_R |\tilde G|^2} 
\sum_{i=1}^{d_R} \bigg| 
\sum_{g\in\tilde G/\tilde S} \bar\gamma^g\, \lp( \omega_i^{g h g^{-1}} \rp)^j
\bigg|^2 
\nonumber\\ &\neq& 0,
\ee

\noindent
as long as $p$ does not divide $j$. Note that in general 
$|F_{h\rightarrow\gamma}|\neq |F_{h^j\rightarrow\gamma}|$. The fact that
was used above is that $|F_{h\rightarrow\gamma}|\neq 0$ implies that
at least two coefficients of different powers of $\omega$ must be different.
Replacing $\omega$ by a power of itself just permutes the coefficients $c_i$
in Eq.~(\ref{eq:omegacoeff}).

The second property is easier to prove in the form of Eq.~(\ref{eq:noS}) 
and states that given $|F_{h\rightarrow\gamma}|\neq 0$ then
\be
\lp|F_{x h x\rightarrow\gamma}\rp|^2 &=& 
\frac{1}{d_R |\tilde G|^2} 
\sum_{i=1}^{d_R} \bigg| 
\sum_{g\in\tilde G} \bar\gamma^g\, \omega_i^{g x h x g^{-1}}
\bigg|^2
\nonumber\\
&=& \frac{1}{d_R |\tilde G|^2} 
\sum_{i=1}^{d_R} \bigg| 
\sum_{g\in\tilde G} \bar\gamma^{g x^{-1}}\, \omega_i^{g h g^{-1}}
\bigg|^2
\nonumber\\
&=& \lp|F_{h\rightarrow\gamma}\rp|^2 \neq 0,
\ee

\noindent
for any $x\in\tilde G$. The second line involves a relabeling of the summation
variable, whereas the third line is true because $\gamma$ is a group 
homomorphism and $\bar \gamma^{-x}$ is just an overall phase.

Together, the two properties imply that, if 
$\lp|F_{h\rightarrow\gamma}\rp|^2$ is non-zero, then so are the amplitudes
$\lp|F_{h'\rightarrow\gamma}\rp|^2$ for any non-trivial $h'=g h^i g^{-1}$.
Unfortunately, even after adding the identity element, this set is in 
general not a group. Furthermore, it remains to be shown that the amplitude
is non-zero for at least one $h$.

\subsubsection{Finding a non-zero amplitude}

It is possible to indirectly show that, for every element $h\in\tilde H$,
there is a pair of representations $R$ and $\gamma$ meeting our requirements,
such that $\lp|F_{h\rightarrow\gamma}\rp|^2\neq 0$.

The basic idea is to consider the regular representation of $\tilde G$. 
Let $\mathcal{H}_{\tilde G}$ be the Hilbert space spanned by the vectors
\be
\ket{g}_{\tilde G}
\ee

\noindent
for $g\in\tilde G$. For the moment, these are just abstract vectors in a
Hilbert space, and therefore we use the above notation to distinguish them
from the anyon magnetic charges. 

The group $\tilde G$ has both a left and a right action on this vector space,
which transforms as the regular representation in both cases.
More generally, we could say that there is an action of the group 
$\tilde G \times \tilde G$ on this vector space given by
\be
\ket{g}_{\tilde G} \longrightarrow \ket{g_1 g g_2^{-1}}_{\tilde G}
\ee

\noindent
for any element $g_1\times g_2\in\tilde G \times \tilde G$.

Let $\mathcal{H}_R$ be the Hilbert space spanned by the vectors of the form
$\ket{M}_R$, where $R$ in an irreducible representation of $\tilde G$. 
These spaces
are also representations of $\tilde G \times \tilde G$ and, in fact, are
irreducible. The space $\mathcal{H}_{\tilde G}$ decomposes
as a sum of irreducible representations of $\tilde G \times \tilde G$
as
\be
\mathcal{H}_{\tilde G} = \bigoplus_{R} \mathcal{H}_R
\ee

\noindent
with each irreducible representation $R$ appearing exactly once.
Fusion corresponds to a further decomposition into the irreducible 
representations of the diagonal group $\tilde G$. 

The state $\ket{I}_{\tilde G}$ transforms as the identity under the
diagonal group, and can therefore be written as a sum of states
$\ket{R(I)}_R$ for different representations $R$. Hence a state 
$\ket{h}_{\tilde G}$ can be written as a sum of states $\ket{R(h)}_R$. If the
state $\ket{h}_{\tilde G}$ has a non-zero projection to a representation
$\gamma$ of the diagonal group, then we know that 
$|F_{h\rightarrow \gamma}|\neq 0$ for at least one irreducible 
representation $R$.

More explicitly, the projection is
\be
P_{\gamma} \ket{h}_{\tilde G} = \frac{1}{| \tilde G |}
\sum_{g\in\tilde G} \bar\gamma^g \ket{g h g^{-1}}_{\tilde G}.
\ee

To make it non-zero, it is sufficient to choose $\gamma$ to be constant
over the stabilizer, $S_h$, in $\tilde G$ of $h$. This is still possible,
even with our requirements that $\gamma$ be one-dimensional and non-trivial,
because $S_h/\tilde H$ is a proper subgroup of the nilpotent
group $\tilde G/\tilde H$. Proper subgroups of nilpotent groups are
always contained in proper normal subgroups because the normalizer
of the proper subgroup is always a larger group (and eventually the operation
of replacing a subgroup with its normalizer must yield a normal subgroup).
This concludes the proof that, for any non-trivial $h\in\tilde H$,
there exists a choice of $\gamma$ and $R$ such that 
$|F_{h\rightarrow \gamma}|\neq 0$.

In fact, for any two non-trivial elements 
$\lambda_1,\lambda_2\in\tilde\Lambda$, the same
representation $\gamma$ is useful because $S_{\lambda_1}=S_{\lambda_2}$. 
However, it is not clear that it is possible to pick $R$ such that both
$|F_{\lambda_1\rightarrow \gamma}|\neq 0$ and 
$|F_{\lambda_2\rightarrow \gamma}|\neq 0$.
This is illustrated by working with the group 
$\Z_5^2\sdp{\theta}(\Z_2\times\Z_3)$, where certain choices of $\gamma$
consistent with the above discussion lead to zero amplitudes for at least one
non-trivial element of $\tilde\Lambda$, no matter which $R$ is used.
On the other hand, the same example does have simultaneous choices of $R$
and $\gamma$ that satisfy all our requirements. It is unclear
to the author whether it is possible, for any group $\tilde G$, to choose $R$ 
and $\gamma$ such that $|F_{\lambda\rightarrow \gamma}|\neq 0$ for all 
non-trivial elements $\lambda\in\tilde\Lambda$ simultaneously.

\subsubsection{\label{sec:bad}Alternative $\tilde\Lambda$}

What happens if $R$ and $\gamma$ cannot be chosen so that 
$|F_{\lambda\rightarrow \gamma}|\neq 0$ over all non-trivial elements 
$\lambda\in\tilde\Lambda$?
While none of the examples in this paper have this problem, if such a case 
arises, we could try to shrink $\tilde\Lambda$. In particular,
if $\tilde\Lambda=\Z_p$, then the problem is solved. That is, because
we can always choose the representations so that the amplitude is non-zero
for some element, and then it is guaranteed to be non-zero for the powers
of that element as well.

The set of functions balanced on $\tilde\Lambda=\Z_p$ can be easily constructed
as simply $\phi(g)=g^i$ for $0<i<p$. However, the probabilistic projection
onto $\tilde\Lambda$ is more difficult. It can be achieved if we are willing to
relax the error model of the probabilistic projections. That is, we 
use an approximate probabilistic projection, where the probabilities and
projected states are close to the desired results. While the results
will be exponentially close in the number of successful fusions, they will
only be polynomially close in the number of actual fusions, and therefore
the machinery of fault tolerant quantum computation must be employed.
Computation with the approximate gate will still
be feasible, but one of the advantages of topological quantum computation, 
that is, the exactness of gates, will be lost.

To construct this approximate projection, consider the amplitude for
the fusion of the electric charges into the vacuum, denoted by 
$F_{h\rightarrow I}$. It is the same quantity that has been dealt with
thus far, only with the representation $\gamma$ replaced by the identity
representation. These quantities have the expression
\be
\lp|F_{h\rightarrow I}\rp|^2 &=&\frac{1}{d_R |\tilde G|^2} 
\sum_{i=1}^{d_R} \bigg| 
\sum_{g\in\tilde G} \omega_i^{g h g^{-1}}
\bigg|^2
\nonumber\\&=&
\frac{1}{d_R |\mathcal{C}_{\tilde G}(h)|^2} 
\sum_{i=1}^{d_R} \bigg| 
\sum_{h'\in\mathcal{C}_{\tilde G}(h)} \omega_i^{h'}
\bigg|^2,
\ee

\noindent
where $\mathcal{C}_{\tilde G}(h)$ is the conjugacy class of $h$ in
$\tilde G$. The amplitudes satisfy the properties
\be
0<\lp|F_{h\rightarrow I}\rp|^2<\lp|F_{I\rightarrow I}\rp|^2
\label{eq:ineq}
\ee

\noindent
for any non-trivial $h\in\tilde H$. The first inequality comes from the
fact that we are summing $p^{th}$ roots of unity and the number of summands
is not divisible by $p$. The second inequality comes from the fact that
$\omega_i^{h'}$ must be non-constant over the conjugacy class of $h$.
The equation
\be
I = \prod_{h'\in\mathcal{C}_{\tilde G}(h)} h',
\ee

\noindent
is true because the right-hand side commutes with all of $\tilde G$,
and therefore must be the identity. Because the number of factors
on the right is not divisible by $p$, $\omega_i$ cannot be constant
over the conjugacy class unless it is the identity. Furthermore, 
since the conjugacy class generates $\tilde H$, and $R$ is non-trivial,
one of the $\omega_i$ must not be the identity. This proves the second 
inequality of Eq.~(\ref{eq:ineq}).

The standard procedure of entangling a state with an electric charge pair,
which is then fused, can then be used. The state is now kept if the
pair fuses into the vacuum, which always has a non-zero probability
of occurring.
The basis state that was entangled with $\ket{R(I)}_R$ will have its amplitude
increased relative to the other basis states. Using braiding to achieve a 
function of the form $f(h) = h a^i$, for some element $a\in\tilde\Lambda$ 
and different values of $i$, we can make the basis states in $\tilde\Lambda$ 
consisting of powers of $a$ have an arbitrarily large amplitude relative 
to the other states. Even if $|F_{h\rightarrow I}|$ varies significantly 
over the non-trivial elements of $\tilde H$, we can use the old 
$\tilde\Lambda$ projector and functions in $\Phi$ to balance out the 
non-trivial elements while increasing the amplitude of the state
with $f(h)= I$. After many repetitions, the basis states with 
flux $a^i b a^{-i}$ can be made to have an amplitude much larger 
then all the other states. This completes
the construction of the approximate probabilistic projection onto the new 
$\tilde\Lambda$ for the special cases when we require $\tilde\Lambda=\Z_p$.

\subsection{Putting it all together}

At this point we have shown the existence of an extended computational space,
with elements labeled by $\tilde H=\Z_p^n$, on which we can perform
the generalized controlled-$X$, and probabilistic projections onto 
$\ket{0,\dots,0}$ and $\ket{\tilde 0,\dots,\tilde 0}$. Furthermore,
there exists a non-trivial subgroup $\tilde\Lambda\subset\tilde H$, such that
we can implement probabilistic projections onto $\tilde\Lambda$ and 
$\ket{0,\dots,0}^\perp\cap\tilde\Lambda$.

To define the real computational subspace, choose a non-trivial element 
$a\in\tilde\Lambda$, and define 
\be
\ket{i}\equiv\ket{a^i b a^{-i}},
\ee

\noindent
for $0\leq i < p$. This subspace corresponds to the subgroup 
$\{a^i\}\subset\tilde\Lambda$ of powers of $a$.

A probabilistic projection onto the real computational space, corresponding
to $\{a^i\}$,
can be achieved in two steps. The first step is to apply the probabilistic
projection onto $\tilde\Lambda$. The second step is repeated for each 
$\lambda\in\tilde\Lambda$ that is not in $\{a^i\}$. For fixed $\lambda$, 
we use an ancilla to conjugate by $\lambda^{-1}$, then do the probabilistic 
projection onto $\ket{0,\dots,0}^\perp\cap\tilde\Lambda$ and then 
conjugate by $\lambda$ using another ancilla:
\be
\sum_{x\in\tilde\Lambda} \alpha_x \ket{x b x^{-1} } 
&\longrightarrow& \sum_{x\in\tilde\Lambda} \alpha_x 
\ket{\lambda x b x^{-1} \lambda^{-1} } 
\nonumber\\
&\longrightarrow& C \sum_{x\in\tilde\Lambda,x\neq\lambda} \alpha_x 
\ket{\lambda x b x^{-1}\lambda^{-1}} \nonumber\\
&\longrightarrow& C \sum_{x\in\tilde\Lambda,x\neq\lambda} \alpha_x 
\ket{x b x^{-1}},
\ee

\noindent
where the probabilistic projection was assumed to succeed in the second step,
and therefore the state is renormalized by the constant $C$. 
The net effect of one such operation is to project out the 
state $\ket{\lambda b\lambda^{-1}}$. If all the projections succeed, then we
will have projected the original state into the computational basis,
completing the probabilistic projection onto $\{a^i\}$.

For the case of qudits with $d=p>2$ we are now done. The generalized
controlled-$X$ behaves as a controlled-$X$ when restricted to act
on the computational space. A probabilistic projection onto $\ket{0}$
is just the probabilistic projection onto $\ket{0,\dots,0}$
because $\ket{0,\dots,0}=\ket{0}$. The probabilistic projection onto
$\ket{\tilde 0,\dots,\tilde 0}$ behaves as a probabilistic projection
onto $\ket{\tilde 0}$ because 
\be
\braket{\tilde i}{\tilde 0,\dots,\tilde 0} \propto \delta_{i,0}
\ee

\noindent
with the caveat that we must use the projection onto the computational
basis to turn the $\ket{\tilde 0,\dots,\tilde 0}$ ancillas into 
$\ket{\tilde 0}$ ancillas. Finally, the probabilistic projection
onto $\ket{0,\dots,0}^\perp\cap\tilde\Lambda$, reduces to a probabilistic
projection onto $\ket{0}^\perp$ when acting on states in the computational
subspace. These are the gates that were proven universal for quantum
computation in Section~\ref{sec:gates}.

\subsubsection{The case $p=2$}

Special treatment must be given to the case when $p=2$, that is, when 
working with qubits. Though all the gates constructed above are valid for
$p=2$, the gate-set is not universal. The problem is that the probabilistic 
projection onto $\ket{0}^\perp=\ket{1}$ does not provide any additional
computational power beyond the probabilistic projection onto $\ket{0}$.

Just as in Section~\ref{sec:magic}, the gate-set can be made universal
given a supply of the magic states:
\be
\ket{\phi_{M1}} &=& \frac{1}{2}
\sum_{i,j} \ket{i}\otimes\ket{j}\otimes\ket{i j},\nonumber\\
\ket{\phi_{M2}} &=& \frac{1}{2}
\sum_{i,j} \omega^{\delta_{i,1}\delta_{j,1}} \ket{i}\otimes\ket{j},
\ee

\noindent
where the second state can be produced from the first one by measuring the
third qudit in the $X$ basis.

The production of the magic state $\ket{\phi_{M1}}$ is the step that 
requires a projection constructed from the fusion of
electric charges. Given our choice of $a\in\tilde\Lambda$ above, 
assume that $b a b^{-1}\in\tilde\Lambda$. This must be the case
if $\tilde\Lambda$ was defined as the intersection of kernels of functions.
Clearly, we can apply a controlled conjugation by $a$, and
therefore, additionally, the controlled conjugation by $b a b^{-1}$ and by
$a b a b^{-1}$. Note that $a\neq b a b^{-1}$ because $b$ was chosen to not 
commute with $a$. 

We begin with the 
$\ket{\tilde 0}\otimes\ket{\tilde 0}\otimes\ket{\tilde 0}$ state
and append a $\ket{0}=\ket{b}$ ancilla. We then conjugate it to obtain
\be
\frac{1}{\sqrt{8}}\sum_{i=0}^{1} \sum_{j=0}^{1} \sum_{k=0}^{1}
\ket{i}\otimes\ket{j}\otimes\ket{k}\otimes\ket{f_{i,j,k}b f^{-1}_{i,j,k}},
\ee

\noindent
where 
\be
f_{i,j,k}=a^{1-i} \lp(b a b^{-1}\rp)^{1-j} x^k,
\ee

\noindent
with $x$ to be determined in a moment. A probabilistic projection
onto $\ket{0,\dots,0}^\perp$ is then applied to the last ancilla, and 
the conjugations are undone.

If the projection succeeds we will have projected out two out of the
initial eight basis states, depending on the value of 
$x\in\{a,b a b^{-1},a b a b^{-1}\}$.
In all cases, the state $\ket{1}\otimes\ket{1}\otimes\ket{0}$ is
removed, and for each of the three values of $x$, one of the other
undesirable basis states is removed. Repeating the procedure
once for each value of $x$ produces the desired magic state $\ket{\phi_{M1}}$. 
Note that the above procedure succeeds because $a^2=1$, and $a$ commutes
with $b a b^{-1}$.

What happens if $b a b^{-1}$ is not in $\tilde\Lambda$? This is the case
when fusions of electric charges into the vacuum must be used. In particular,
instead of projecting out the undesirable basis states, we increase the
amplitude of the desired basis states, and obtain an ancilla that is 
exponentially close to the desired magic state. The procedure
is almost unchanged, except that the function involved is
\be
f_{i,j,k}=a^{i} \lp(b a b^{-1}\rp)^{j} x^{1-k},
\ee

\noindent
and the function $f_{i,j,k}'=a^{i} \lp(b a b^{-1}\rp)^{j}$ must also be
used to adjust the relative amplitude of $\ket{1}\otimes\ket{1}\otimes\ket{1}$
with respect to the other desired states.

In either case, we have now shown that case of qubits can be dealt with in a 
similar fashion to the general qudit case, and therefore, we have completed 
the construction of universal quantum computation for anyons based on
solvable non-nilpotent groups.

\section{\label{sec:leak}Leakage Correction}

Before concluding this paper, it is important to address the issue of
fault tolerance. A physical system with anyons will have sources of errors
due to the finite separation of anyons and non-zero temperature (see
Refs.~\cite{me2002, Ogburn:1998, Preskill:1997uk} for details). While
the probability of error is exponentially small in the distance and 
temperature, it is in general non-zero. These errors could be especially 
relevant if anyons are used as long term quantum memory, in which case error 
correcting codes must be employed.

While most of the machinery of error correcting codes can be applied directly
to anyons, it requires that states with errors remain within the 
computational subspace (that is, the subspace on which universal quantum 
computation can be done). For our model of computation, this is only a
small subspace corresponding to anyons that are
magnetic charges with fluxes such as $a^i b a^{-i}$, and arranged in pairs of 
trivial total flux. Note that only the magnetic charges need error correction
as they are the ones in which the quantum state is stored.

All that is required to perform quantum error correction is to be able
to replace qudits that have ``leaked out'' of the computational subspace
with arbitrary states that are in the computational subspace. This step can
then be followed by the standard error correcting step, which will remove 
the errors. The leakage correction step is equivalent to the 
swap-if-leaked gate described by Kempe \textit{et al}. \cite{Kempe:2001}.

In Ref.~\cite{me2002} a leakage correction scheme was presented for 
non-solvable anyons. While a similar scheme could be constructed for the 
solvable anyons discussed in the present paper, it will be easier to 
present a generic leakage correction scheme that can also be applied to
anyons.

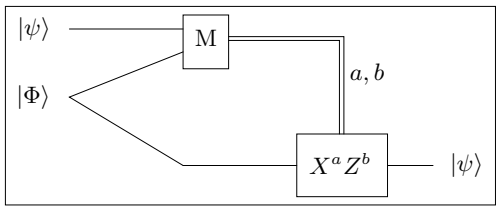
\begin{figure}
\setlength{\unitlength}{0.0002in}
{
\begin{picture}(12948,5259)(-1100,-210)
\path(-1100,-210)(11848,-210)(11848,5049)(-1100,5049)(-1100,-210)
\path(600,4447)(3600,4447)
\path(3600,3847)(600,2647)(3600,847)
\path(3600,847)(6600,847)
\path(6600,1672)(9000,1672)(9000,22)
	(6600,22)(6600,1672)
\path(3600,4822)(4800,4822)(4800,3397)
	(3600,3397)(3600,4822)
\path(4800,4150)(7725,4150)(7725,1672)
\path(4800,4250)(7850,4250)(7850,1672)
\path(9000,847)(10200,847)
\put(100,4447){\makebox(0,0)[rc]{$|\psi\rangle$}}
\put(100,2647){\makebox(0,0)[rc]{$|\Phi\rangle$}}
\put(4200,4200){\makebox(0,0)[cc]{M}}
\put(7750,900){\makebox(0,0)[cc]{$X^a Z^b$}}
\put(10650,847){\makebox(0,0)[lc]{$|\psi\rangle$}}
\put(8000,3247){\makebox(0,0)[lc]{$a,b$}}
\end{picture}
}
\caption{\label{fig:unleak}Leakage correction circuit}
\end{figure}

The scheme is simply to teleport a computational qudit to a fresh qudit.
The standard steps, shown in Figure~\ref{fig:unleak}, are first to 
create the entangled ancilla 
$\ket{\Phi}=\sum_i \ket{i}\otimes\ket{i}/\sqrt{d}$, and then
measure the computational qudit and the first ancilla qudit in the basis
$\ket{a,b}=X^a Z^b\otimes I \ket{\Phi}$, obtaining outcome $a,b$. The
correction gate $X^a Z^b$ is then applied to the second ancilla qudit,
which now becomes part of the computational space. All these operations
can be performed using the anyon gates discussed so far. 

If the original qudit was in the computational space, then its state
will be flawlessly transfered into the new qudit (in our case, a
fresh anyon pair). However, if the original qudit had leaked, then the new
qudit will be guaranteed to be in the computational subspace, because it was
obtained by applying Pauli operators to a qudit known to be in the 
computational subspace. This is the desired leakage correction protocol.

In fact, this scheme can be applied to almost any system, as long as we can
guarantee that the measurement of the first two qudits will not affect the 
third qudit in any way, as should be the case if they are sufficiently
separated.

The leakage correction scheme has caveat from a theoretical standpoint, though.
We are effectively
assuming that we possess a classical leakage detection machine, through
which the data ``$a,b$'' is run. That is, if the measurement produced an
outcome in the form of a voltage, and then the gate $X^a Z^b$ was constructed
as a Hamiltonian controlled by this voltage, we would need to guarantee
that only the $d^2$ acceptable voltage signals could reach the machine 
operating on the third qudit. However, in practice, leakage correcting a 
classical signal is trivial, as classical information can be measured 
without any negative side effects.

A very similar scheme can be produced given a quantum system
that is known to have exactly $d$ states. The qudit is simply swapped
into the new system, the first system is then erased and restored into the
computational space, and then the qudit is swapped back. In this context,
the teleportation scheme is in effect a way of swapping a qudit into a 
classical system.

Though the leakage correction scheme was discussed in general terms, it clearly
applies to the anyons discussed in this paper, and its use allows quantum
error correction and fault tolerance to be employed. We have therefore
shown that even in the presence of small sources of noise, the anyons can
still be used for universal quantum computation.

\section{Concluding Remarks}

The main result of this paper is that anyons from finite groups that are
solvable
but not nilpotent are capable of universal quantum computation. This set
includes many groups of small size, which are more likely to be found in
a physical system. Combined with the results of Ref.~\cite{me2002},
we have proven that every finite group that is not nilpotent produces anyons
capable of universal quantum computation. 

Furthermore, except for the groups where the methods of Section~\ref{sec:bad} 
must be used, the computations with anyons can be made error free in the 
following sense:
in the theoretical limit of zero temperature and infinite separation between 
anyons, an arbitrarily long calculation can proceed without the need of
error correction. 
The elementary unitaries are always perfect, whereas the measurements
are either perfect, or are known to have failed (i.e., when none of
the probabilistic projections succeed). This occurs with a probability 
that can be made exponentially small in the number of fusions. 
Of course, a real system will have additional exponentially small errors 
due to finite size and temperature effects.

The physical requirements for the constructions in this paper include a
supply of electric charge ancillas, in addition to the requirements of 
Ref.~\cite{me2002}. The necessity of the electric charges may present
an extra source of difficulties for a real implementation.
The exception is $S_3$, in which case only magnetic charges are 
required, as mentioned at the end of Section~\ref{sec:sdp}. 
In either case, the issue of producing the elementary 
electric or magnetic ancillas is not addressed in this paper, though a 
generalization of the construction in Ref.~\cite{me2002} may be sufficient.

Another open question is whether anyons from non-abelian 
nilpotent groups are capable of universal quantum computation. Additionally,
not much is known about computing with anyons that do not belong to the 
electric and magnetic charge model discussed in this paper. 
On the other hand, 
the universality of anyons from certain continuous groups has been 
discussed in Refs.~\cite{Freedman:2000,Freedman:2001}.

Of course, the most important open question is whether we can find a
laboratory system with anyons out of which a quantum computer 
can be built. The requirement of a two-dimensional space severely limits the
possibilities. However, certain exotic systems such as the fractional levels
of the quantum Hall effect may contain non-abelian anyons. Another 
option is the possibility of engineering a system with the desired anyons.
Recent proposals include using optical lattices \cite{Duan:2002} or
Josephson-junction arrays \cite{Doucot:2003mh}. In the latter case, an explicit
array is constructed that simulates $S_3$ gauge theory on a lattice. Ideally,
one day such a system could be used to turn the ideas presented
here into a working quantum computer.

\begin{acknowledgments}

Much of this work is inspired by the construction of Alexei Kitaev, who
showed that universal computation was possible with anyons based on
the group $S_3$. His construction used the magic state
\be
\frac{1}{\sqrt{3}}\lp(\ket{0} - \ket{1} - \ket{2}\rp)
\ee

\noindent
to build a qubit Toffoli gate. Many of the above ideas were captured by the
unpublished notes of John Preskill. I am also indebted to Charlene Ahn and
Ben Toner who were kind enough to read and review this paper.

This work was supported in part by
the National Science Foundation under grant number EIA-0086038 and
by the Department of Energy under grant number DE-FG03-92-ER40701.

\end{acknowledgments}


\end{document}